\documentclass[aps,prb,twocolumn,showpacs,superscriptaddress]{revtex4-2}  
\usepackage{graphicx}  
\usepackage{dcolumn}   
\usepackage{bm}        
\usepackage{amsmath}
\usepackage{amsthm}
\usepackage{amssymb}
\usepackage{bbold} 
\usepackage{mathrsfs}
\usepackage{array}
\usepackage[dvipsnames,usenames]{color}
\usepackage{soul} 
\usepackage{ulem} 
\usepackage[bookmarks=true,colorlinks,linkcolor=blue,urlcolor=blue,citecolor=blue]{hyperref}

\newcommand{\be}{\begin{equation}}
\newcommand{\ee}{\end{equation}}
\newcommand{\bea}{\begin{eqnarray}}
\newcommand{\eea}{\end{eqnarray}}
\newcommand{\mc}{\mathcal}
\newcommand{\mb}{\mathbf}

\usepackage{physics}

\begin{document}

\normalem

\title{Electronic interactions in a
vacancy-engineered honeycomb lattice: transition from a nodal-line semimetal to a magnetic insulator}
\author{Andressa R. Medeiros-Silva}
\affiliation{Instituto de F\'isica, Universidade Federal do Rio de Janeiro, Rio de Janeiro, RJ 21941-972, Brazil}
\affiliation{Department of Physics, University of Houston, Houston, Texas, 77004, USA}
\author{Mariana Malard}
\affiliation{Centro Internacional de F\'isica and Faculdade UnB Planaltina, Universidade de Bras\'ilia, Brasília, DF, 70904-910, Brazil}
\author{Rodrigo G. Pereira}\affiliation{International Institute of Physics and Departamento de F\'isica, Universidade Federal do Rio Grande do Norte, Natal, RN, 59078-970, Brazil}
\author{Thereza  Paiva}
\affiliation{Instituto de F\'isica, Universidade Federal do Rio de Janeiro, Rio de Janeiro, RJ 21941-972, Brazil}

\begin{abstract}
Nodal-line semimetals (NLSMs) harbor a variety of novel physical properties owing to the particularities of the band degeneracies that characterize the spectrum of these materials. In symmetry-enforced NLSMs, band degeneracies, being imposed by symmetries, are robust to arbitrarily strong perturbations that preserve the symmetries. We investigate the effects of electron-electron interactions on a recently proposed vacancy-engineered NLSM known as holey graphene. Using mean-field calculations and quantum Monte Carlo simulation, we show that the Hubbard model on the depleted holey-graphene lattice at half-filling exhibits a transition from a NLSM to an insulating antiferromagnetic phase for an arbitrarily weak repulsive interaction $U$. In contrast to the semi-metal-insulator transition in the pristine honeycomb lattice, which occurs at a finite critical value of $U$, in the depleted lattice, the transition at $U=0$ is associated with a van Hove singularity arising from the crossing of accidental nodal lines and those enforced by symmetry. We also employ linear spin wave theory (LSWT) to the effective Heisenberg model in the strong-coupling limit and obtain the global antiferromagnetic order parameter $m_{\rm AFM} \approx 0.146$. The order parameters from both QMC and LSWT agree quantitatively. Our findings indicate that vacancy engineering offers an effective way to tailor the magnetic properties of quantum materials. 
\end{abstract}


\maketitle

\section{Introduction}

As primal as it is, dating back to the beginnings of quantum mechanics~\cite{Neumann1929}, the study of band degeneracies in crystals continues to yield interesting physics. Semimetals are materials whose band degeneracies occur at points -- nodal point semimetals -- or along lines -- nodal line semimetals (NLSMs) -- in the Brillouin zone of otherwise gapped bands~\cite{Murakami2007,Burkov2016}. The lower dimensionality of the Fermi surface of semimetals underpins a wide range of phenomena: from fundamental aspects involving topological charges~\cite{Weng2015}, Fermi arcs~\cite{Wan2011}, and chiral anomaly~\cite{Zyuzin2012}, to technological applications, notably for topological quantum computing~\cite{Burkov2016} and for designs of quantum devices based on proximity effects~\cite{Zutic2019}.

Band degeneracies fall into two categories. A \emph{symmetry-enforced} band degeneracy owes its existence solely to some set of crystal symmetries which must include at least one nonsymmorphic symmetry ~\cite{MichelZak1999}. 
  \emph{Accidental} band degeneracies of one- or two-dimensional crystals require a combination of symmetries and tuning a microscopic parameter of the material. In three dimensions, the availability of sufficiently many tunable parameters (three momentum coordinates and one material parameter) leads to accidental band degeneracies even without symmetries ~\cite{Murakami2007}, the nodes of Weyl semimetals being the most prominent instance.  

Symmetry-enforced band degeneracies have been intensively investigated in the field of topological semimetals ~\cite{Zhao2016,Schnyder2018,Young2015}. Symmetry-enforced NLSMs, in particular, have been predicted in various three-dimensional (3D) groups~\cite{Zhang2018,ParkL2021}. We note that the presence of nonsymmorphic symmetries in the compounds falling in those groups is related to the presence of more than one type of atom in their unit cell. 

Coming to 2D materials, hexagonal lattices~\cite{Xia2019}, honeycomb-Kagome lattices~\cite{Lu2017}, and PbFCl-type structures~\cite{BinGao2019} have been reported to realize accidental NLSMs. These proposals suffer from accidental nodal lines being particularly fragile to perturbations in 2D, especially to spin-orbit coupling (SOC), which essentially prevents their utility in proximity-effect based devices. 
As an alternative to solid-state realizations, NLSMs have recently been reported in experiments of ultracold fermions in optical lattices \cite{Song19}, which offer an unprecedented possibility of engineering nodal lines and fine-tune correlations. 

In recent papers authored by one of us and collaborators \cite{Malard21,Malard22_PRB,Malard22_SciRep}, the following idea was advanced: Vacancy engineering can produce a 2D monoatomic, and yet symmetry-enforced, NLSM by turning a symmorphic crystal into a nonsymmorphic one. Quite surprising at first, besides the targeted symmetry-enforced nodal lines, the proposed materials also exhibit accidental nodal lines that resist very strong SOC, with strength even exceeding the experimental values available then.

Despite the significant effort to understand the topological properties of semimetals, the role of electron-electron (\textit{e-e}) interactions in these materials is less clear. Electronic correlations on NLSMs have been recently probed in ZrSiSe, showing a reduction of the Fermi velocity and a strong suppression of the Drude spectral weight due to short-range interaction effects \cite{Shao20}. Similar results have been obtained from angle-resolved photoelectron spectroscopy, which revealed a light-induced quasiparticle renormalization due to strong \textit{e-e} interactions\,\cite{Gatti2020}. Other NLSM candidates also exhibit correlated properties, such as magnetism in HoSbTe\,\cite{Yang2020} and \textit{A}$_{2}$B$_{3}$ (\textit{A} = Ti-Ni)\,\cite{Sun2022}, or superconductivity\,\cite{Shang2022}. Theoretically, \textit{e-e} interactions in NLSMs have been examined in connection to the emergence of phases with broken symmetry\,\cite{Koshino2016,Roy2017,Wang2017,Jose2020,Wu2023} and as corrections to physical quantities\,\cite{Huh2016,Segovia2020}.

\begin{figure}[t]
\centering
\includegraphics[width=0.98\linewidth]{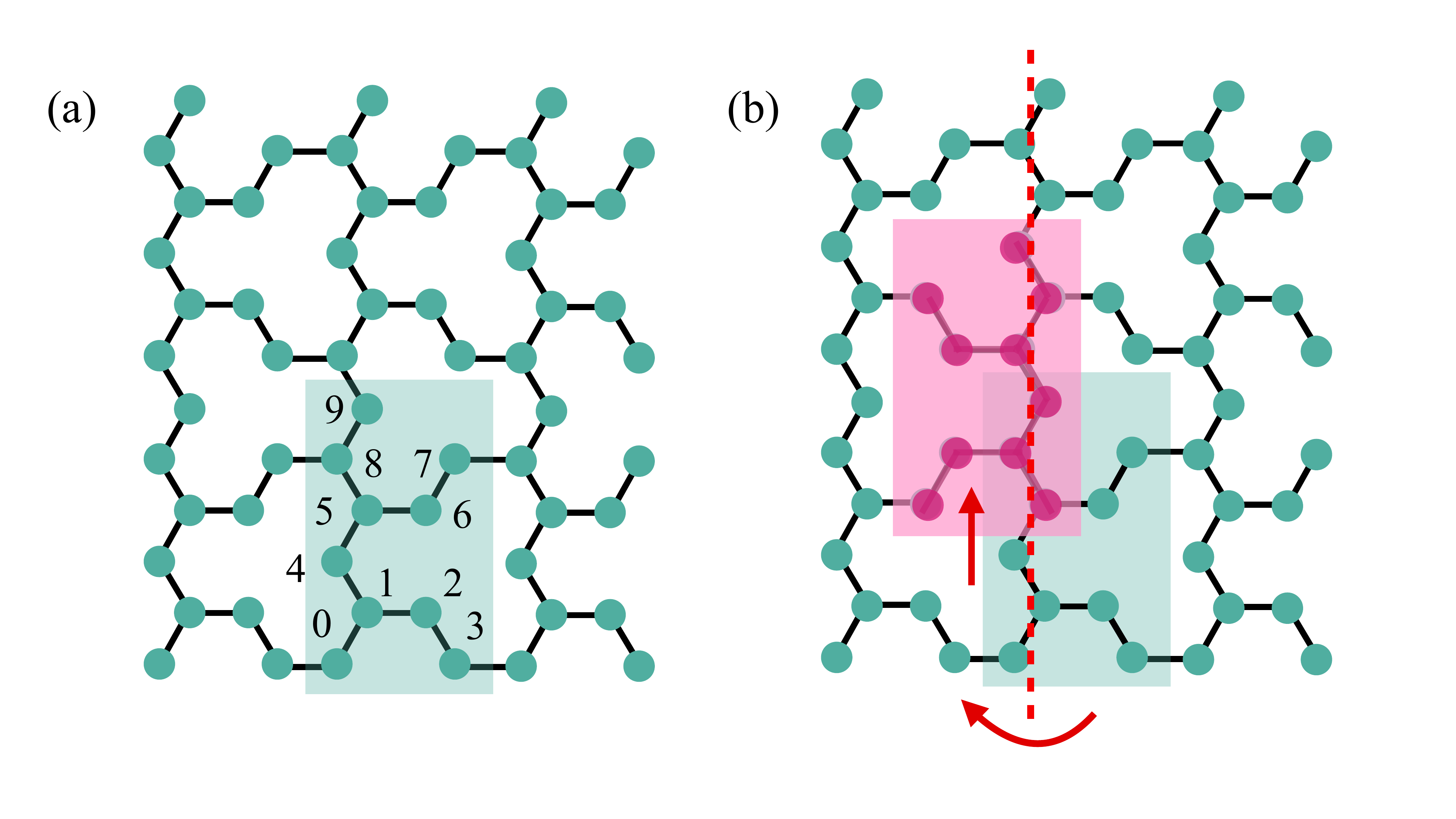} 
\caption{(a)\;The honeycomb lattice with depletions. The highlighted rectangle is the unit cell and the basis sites are numbered. (b)\;The glide plane is depicted by the red dashed line.}
\label{fig:honeydec}
\end{figure}

Back to Refs. \cite{Malard21,Malard22_PRB,Malard22_SciRep}, one of the structures investigated therein is obtained by depleting $1/5$ of the sites of a honeycomb lattice, as depicted in Fig.\,\ref{fig:honeydec}\,(a), wielding a nonsymmorphic glide-plane symmetry, see Fig.\,\ref{fig:honeydec}\,(b).
Glide-plane, inversion, and time-reversal symmetries enforce nodal lines in the BZ of the depleted structure\,\cite{Malard22_SciRep}. For short, let us call the so-obtained 2D vacancy-engineered symmetry-enforced NLSM a \emph{nonsymmorphic holey graphene}. We note that the proposed mechanism does not rely on chemical composition. Hence, Fig.\,\ref{fig:honeydec}\,(a) could, in principle, represent any material that realizes a honeycomb lattice in its vacancy-free form, such as silicene, germanene, and borophene, besides graphene. For graphene, similar lattice-engineering methods have been reported using defects \cite{Kotov2012} and by manipulating single atoms on a substrate \cite{Slot2017}.

In this work, we employ a mean-field approach and an unbiased determinatal quantum Monte Carlo (DQMC) simulation to investigate \textit{e-e} interactions in the nonsymmorphic holey graphene of Fig.\,\ref{fig:honeydec}\,(a). We are particularly interested in the possibility that interactions drive symmetry-breaking instabilities, gapping out the nodal lines. The codimension of the nodal line plays a crucial role in this analysis. In three dimensions, nodal lines (codimension $p=2$) lead to a vanishing density of states  \cite{Burkov11}, which implies that short-range interactions are perturbatively irrelevant in the renormalization-group sense. On the other hand, a nodal line in two dimensions  (codimension $p=1$) generates a finite density of states. In the particular case of the model for nonsymmorphic holey graphene with uniform hopping parameters, we find that the spectrum exhibits a crossing between a symmetry-enforced and an accidental nodal line, which gives rise to a van Hove singularity at the Fermi level. We will show that this effect leads to an instability towards an antiferromagnetically ordered state for arbitrarily weak \textit{e-e} interactions. By contrast, for the Dirac semimetal in the half-filled pristine honeycomb lattice, the transition to an antiferromagnetic (AFM) Mott insulator requires a finite interaction strength $U_{c}=3.8$ (in units of the hopping amplitude) \cite{Paiva05,Sorella12,Assaad2013}.

The paper is structured as follows:
In Sec.\,\ref{sec:HQMC}, we present and discuss the Hubbard Hamiltonian and the methodologies employed in this work. Our mean-field and DQMC results are presented in Sec.\,\ref{sec:results}, as well as a linear spin wave approach in the strong coupling limit.
The main conclusions and future perspectives derived from this work are relayed in Sec.\,\ref{sec:conc}.


\section{Model and Methodology}
\label{sec:HQMC}

We investigate the properties of interacting fermions on a nodal-line semimetal system created by engineering the lattice presented in Fig.\,\ref{fig:honeydec}\,(a).
In order to facilitate the following discussions, we assume single $s$-orbitals with a Hubbard on-site interaction, while keeping the hopping integrals only between nearest neighbor (NN) sites.
Therefore, the Hamiltonian reads
 \begin{align}\label{Eq:Hamilt}
 \nonumber \mathcal{H} = & -\sum_{\substack{\langle \textbf{i},\textbf{j} \rangle},\sigma} t_{\mathbf{i},\mathbf{j}} \big( c_{\textbf{i} \sigma}^{\dagger}c_{\textbf{j} \sigma}+ {\rm H.c.} \big) - \mu \sum_{\substack{\textbf{i}}, \sigma} n_{\textbf{i},\sigma}
 \\  & + U   \sum_{\substack{\textbf{i}}} \big(n_{\textbf{i},\uparrow} - 1/2 \big) \big(n_{\textbf{i},\downarrow} - 1/2\big),
 \end{align}
with the sums running over all lattice sites, and $\langle \mathbf{i}, \mathbf{j} \rangle$ denoting NN sites. 
Here, we define the notation $\mathbf{i} = \boldsymbol\alpha + \mathbf{r}$, with $\boldsymbol\alpha$ denoting the position of the $\alpha$-th orbital  ($\alpha=0,\dots,9$), and $\mathbf{r}$ the position of the unit cell.
We use the second quantization formalism, in which $c^{\dagger}_{\mathbf{i} \sigma}$ ($c^{\phantom{\dagger}}_{\mathbf{i} \sigma}$) describe creation (annihilation) operators of electrons on a given site $\mathbf{i}$ and spin $\sigma$, while $n_{\mathbf{i}\sigma} \equiv c^{\dagger}_{\mathbf{i} \sigma} c_{\mathbf{i} \sigma}$ are number operators.
The first term on the right-hand side of Eq.\,\eqref{Eq:Hamilt} denotes the electronic hopping of amplitude $t$, while the second one corresponds to their chemical potential $\mu$.
The last term describes the on-site repulsive interaction of strength $U$, which we assume to be uniform. To preserve the glide symmetry and, consequently, the nodal lines, one must impose that the following pairs of orbitals are equivalent (see Fig.\,\ref{fig:honeydec}): ($s0$,$s5$), ($s1$,$s8$), ($s2$,$s7$), ($s3$,$s6$), and ($s4$,$s9$). Similarly, the equivalent hopping integrals $t_{\alpha\alpha'}$ are: $t_{01}=t_{58}$, $t_{12}=t_{78}$, $t_{23}=t_{67}$, $t_{14}=t_{89}$, $t_{45}=t_{90}$, and $t_{56
}=t_{03}$.  If we impose spatial inversion in addition to glide symmetry, we are left with three sets of equivalent orbitals: ($s0$,$s1$,$s5$,$s8$), ($s2$,$s3$,$s6$,$s7$), and ($s4$,$s9$).  Hereafter, unless otherwise mentioned, we set $t_{\mathbf{i},\mathbf{j}}=t$, while defining the lattice constant as unity.

We investigate the properties of this Hamiltonian using a mean-field approach and unbiased determinant quantum Monte Carlo (DQMC) simulations\;\cite{Blankenbecler81,Hirsch85,White89}. We also use linear spin-wave theory (LSWT) to probe the ground state magnetic properties of the strong-coupling limit, $U \to \infty$, as the system can be mapped onto the Heisenberg model at half-filling.

DQMC maps an interacting system onto a non-interacting one by introducing bosonic auxiliary fields.
Briefly, the kinetic (quadratic) and potential (quartic) terms in the partition function are separated using a Trotter-Suzuki decomposition, with the inverse temperature $\beta = L / \Delta \tau$ corresponding to the imaginary time coordinate, with $\Delta \tau$ being the discretization parameter, and $L$ the number of time slices. This decomposition has an error proportional to $\Delta \tau^2$ and becomes exact in the limit $\Delta \tau \rightarrow 0$. In our simulations, $\Delta \tau \leq 0.1$, ensuring that the Trotter-Suzuki error is smaller than the statistical error from Monte Carlo sampling. We then transform the many-particle quartic operators into single-particle quadratic ones using a Hubbard-Stratonovich transformation, which introduces bosonic auxiliary fields sampled by the Monte Carlo method. This procedure allows us to compute the Green's functions and, through Wick's decomposition, all higher-order correlation functions. More details about the methodology can be found in the Refs.\,\cite{assaad02,Santos03,gubernatis16,Becca17}.

On the other hand, the mean-field approach performs an approximation in the interaction term by breaking the symmetry of the Hamiltonian.
For the Hubbard model, the \textit{SU(2)} symmetry can be broken along the \textit{z}-component of the spin, so the interacting term may be approximated as
\begin{align}\label{Eq:Inter_MF}
\nonumber  U n_{\mathbf{i},\uparrow} n_{\mathbf{i},\downarrow} \approx\;
& U \sum_{\sigma} \left[ \frac{\langle n_{\mathbf{i}} \rangle}{2} - \sigma \langle m_{\mathbf{i}} \rangle \right]c^{\dagger}_{\mathbf{i} \sigma} c_{\mathbf{i} \sigma}  \\
& - U \left[\frac{\langle n_{\mathbf{i}} \rangle^2}{4} - \langle m_{\mathbf{i}} \rangle^2 \right]
\end{align}
with $\langle n_{\mathbf{i}} \rangle = \langle c^{\dagger}_{\mathbf{i} \uparrow} c_{\mathbf{i} \uparrow} + c^{\dagger}_{\mathbf{i} \downarrow} c_{\mathbf{i} \downarrow} \rangle$ being the average number of electrons on site $\mathbf{i}$, while $\langle m_{\mathbf{i}} \rangle = \langle S_{\mathbf{i}}^{z} \rangle = \frac{1}{2} \langle c^{\dagger}_{\mathbf{i} \uparrow} c_{\mathbf{i} \uparrow} - c^{\dagger}_{\mathbf{i} \downarrow} c_{\mathbf{i} \downarrow} \rangle$ its magnetization.
Since the lattice is bipartite, we assume an AFM  state in which the sign of $\langle m_{\mathbf{i}} \rangle$ alternates between even and odd sublattices (i.e., between even and odd values of $\alpha$), in a way that the spatial dependence of $\langle n_{\mathbf{i}} \rangle$ and $\langle m_{\mathbf{i}} \rangle$ preserves translational symmetry. Further details of our mean-field approach are discussed later.

\begin{figure}[t]
\centering
\includegraphics[width=0.98\linewidth]{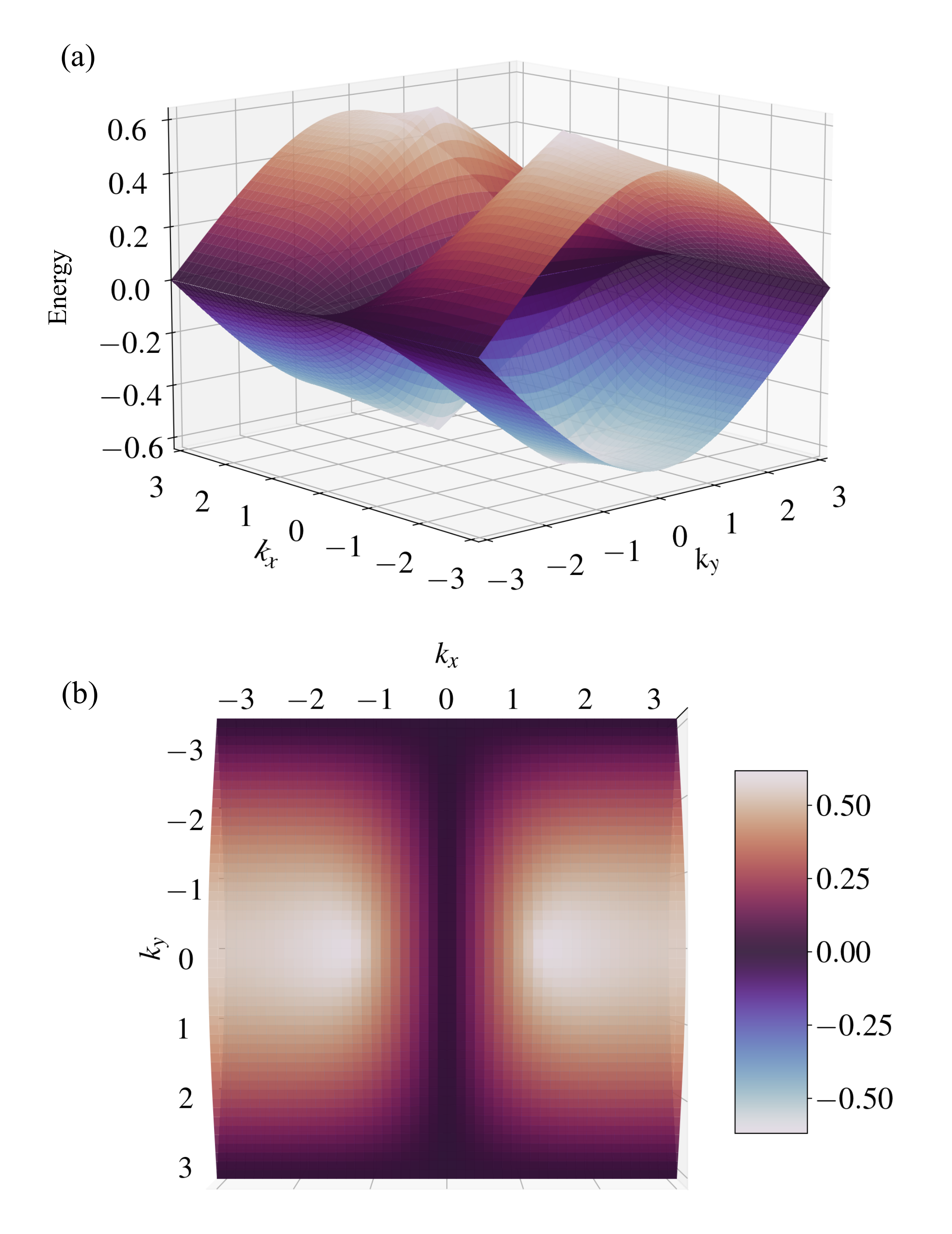} 
\caption{(a)\;Side and (b)\;top view of the dispersion relation focusing on the two bands closer to the Fermi level ($E = 0$) at half filling.
The bands cross at the Fermi level (darker color), forming the nodal lines along $k_x = 0$ and $k_y = \pi$. The energy scale is expressed in units of $t = 1$.}
\label{fig:fermi_surface}
\end{figure}

\section{Results}
\label{sec:results}

\subsection{Vacancy-engineered nodal-line semimetal: the non-interacting limit}

As discussed earlier, a NLSM may be created by engineering a lattice with periodic depletions that give rise to glide, point inversion, and time reversal symmetries\;\cite{Malard22_SciRep}. For example, it is achieved by engineering a honeycomb lattice depleted with $1/5$, as shown in Fig.\,\ref{fig:honeydec}\,(a), with the shaded area defining the unit cell, while the numbers are the site labels.
As shown in Fig.\,\ref{fig:honeydec}\,(b), this structure has a glide symmetry, with the vertical dashed (red) line  representing the glide plane. In this subsection, we discuss the non-interacting properties of this NLSM, i.e., we set $U=0$, leaving the interaction effects to the next subsections.

The Fourier transform of the non-interacting Hamiltonian leads to a block-diagonal $10\times 10$ matrix that can be diagonalized numerically.
Due to the particle-hole symmetry, the nodal lines appear at zero energy, coinciding with the Fermi level at half filling.
Figure \ref{fig:fermi_surface}~(a) displays the two bands that cross the Fermi level and touch along the nodal lines presented in Fig.\,\ref{fig:fermi_surface}\,(b).
Interestingly, there are two nodal lines: one at $k_y = \pi$ and another at $k_x = 0$.
As discussed in Refs.\,\cite{Malard21,Malard22_PRB,Malard22_SciRep}, the former is symmetry-enforced, while the latter is accidental.

\begin{figure}[t] 
\centering
\includegraphics[width=0.98\linewidth]{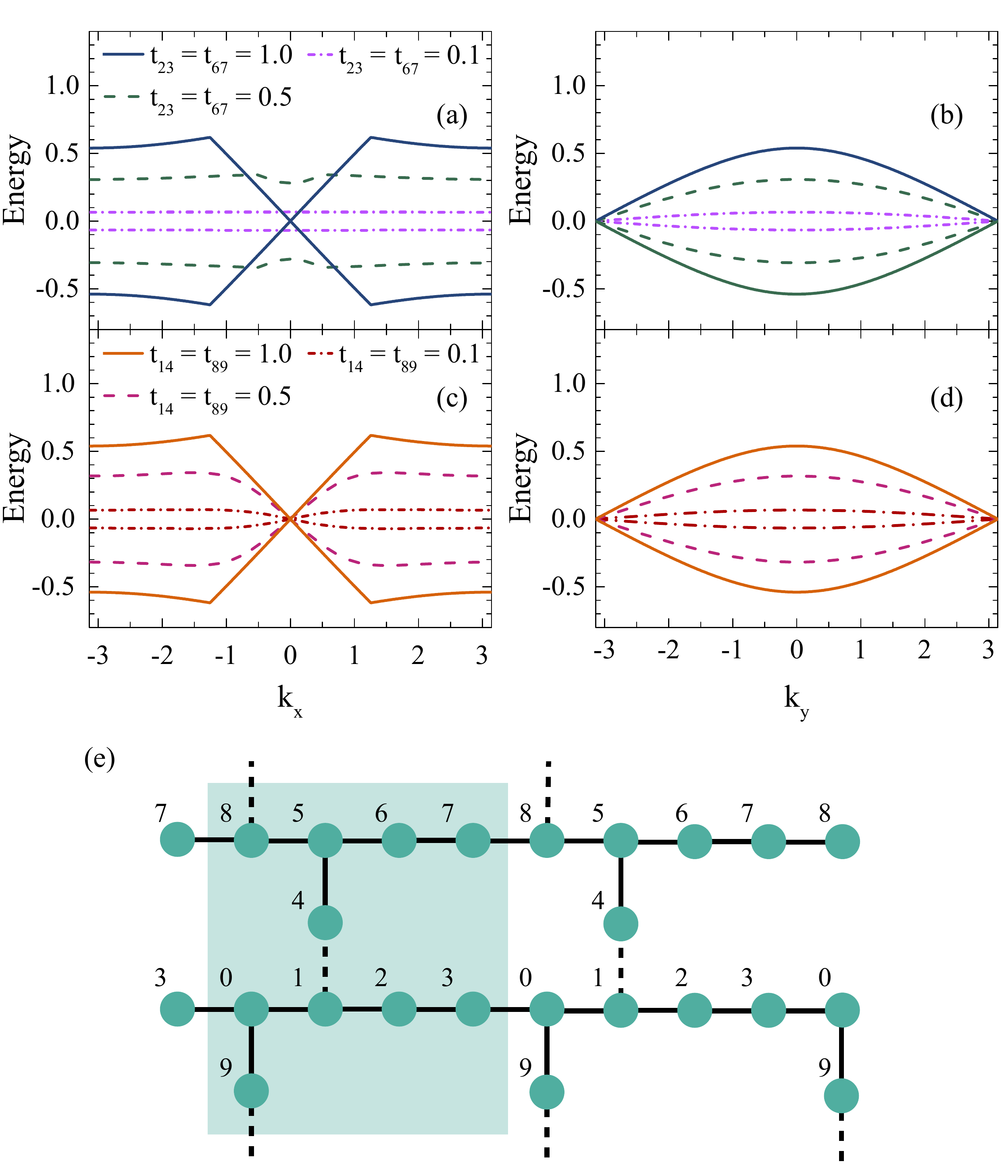}
\caption{Energy dispersion for fixed $k_y = 0$ [panels (a) and (c)] and $k_x = \pi$ [panels (b) and (d)]. 
The nodal-line at $k_x = 0$  is gapped out whenever $t_{23}=t_{67}\neq1.0$, but remains unaffected when $t_{14}=t_{89}\neq1.0$, where $t=1.0$ is the value of the hopping parameter for the other glide-related pairs. Expectedly, the nodal-line at $k_y = \pi$ is not destroyed for  when changing the value of any of the glide-related pairs. (e) Different arrangement of the depleted honeycomb lattice, which shows the decoupling into two separate lattices when $t_{14} = t_{89} \rightarrow 0$ (dashed lines).}
\label{fig:NLvarying}
\end{figure}

Although the results presented in Fig.\,\ref{fig:fermi_surface} were obtained by fixing $t_{\mathbf{i},\mathbf{j}} = t$, we can test the stability of the nodal lines by  considering inhomogeneous hopping while preserving the glide symmetry.
For example, when we set $t_{23}=t_{67} \neq t$ [See Figs. 3(a)-(b)], the (horizontal) nodal line at $k_y = \pi$ is maintained, while the (vertical) nodal line at $k_x = 0$ is destroyed. 
We illustrate  this feature by fixing $k_y = 0$ and varying $k_x$, or by fixing $k_x = \pi$ and varying $k_y$, as shown in Figs.\,\ref{fig:NLvarying}\,(a) and (b), respectively. 
In fact, any difference among the pairs of hopping integrals ($t_{01}$,$t_{58}$), ($t_{12}$,$t_{78}$), ($t_{23}$,$t_{67}$) and ($t_{56}$,$t_{03}$) is enough to gap out the accidental nodal line at $k_x = 0$ while the symmetry-enforced one at $k_{y}=\pi$ is unremovable.

Interestingly, unlike the previous case, changing the values of  ($t_{14}$,$t_{89}$) and ($t_{45}$,$t_{90}$) does not destroy the accidental nodal line at $k_x = 0$.
This behavior is shown in Fig.\,\ref{fig:NLvarying}\,(c)-(d)for the case of changing ($t_{14}$,$t_{89}$), where one may notice the presence of both nodal lines for different values of ($t_{14}$,$t_{89}$), with the crossing bands collapsing into a twofold degenerate flat band when $t_{14}=t_{89}\to 0$.
To understand the formation of these flat bands, we note that for  $t_{14}=t_{89}=0$ the system reduces to a set of decoupled chains, as illustrated in Fig.\,\ref{fig:NLvarying}\,(e). The fact that   the chains are bipartite and have an odd number of sites per unit cell implies compact localized states, thus flat bands\,\cite{Ramachandran2017,Lieb89}. Therefore, we expect that in the half-filled ground state, the limiting case $t_{14}=t_{89}\to 0$ would favor magnetism for any $U/t > 0$, as provided by Lieb's theorem\,\cite{Lieb89}.

\begin{figure}[t]
\centering
\includegraphics[width=0.98\linewidth]{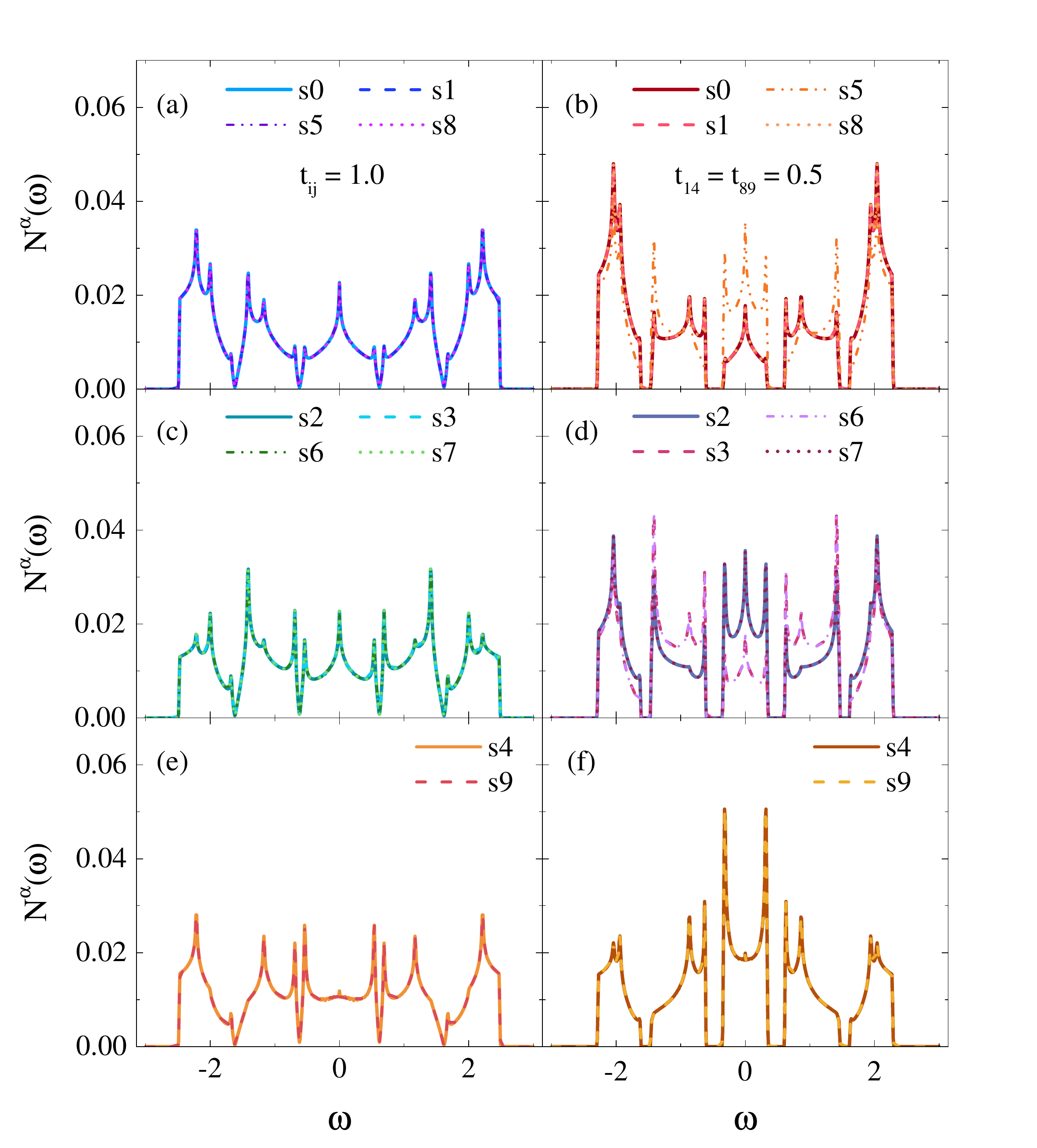} 
\caption{Local density of states for orbitals sites $s0$, $s5$, $s1$ and $s8$ (a)-(b), $s2$, $s7$, $s3$ and $s6$ (c)-(d), and $s4$ and $s9$ (e)-(f). The panels on the left display the LDOS for homogeneous hoppings ($t_{ij} = 1.0$) and the panels on the right display data for $t_{14}=t_{89} = 0.5$.
The van Hove singularities present in the homogeneous case [panels (a) and (c)] persist when hopping anisotropy in $t_{14}=t_{89}$ is introduced [panels (b) and (d)]. For $s4$ and $s9$, the van Hove singularity at $\omega=0$ is weak in both the isotropic and anisotropic scenarios [panels (e) and (f).]}
\label{fig:dos_hn}
\end{figure}

Examining the local density of states (LDOS) in the noninteracting case is also important.
Taking the time Fourier transform and performing an analytical continuation of the single-particle Green's function (see, e.g.\,Refs.\ \cite{Nandini2013,Beach1999} for details), we obtain the spectral functions
\begin{equation}
A^\alpha_\sigma(\mathbf{k},\omega)=-\dfrac{1}{\pi}\operatorname{Im} G^\alpha_\sigma(\mathbf{k},\omega+ i 0^+),
\end{equation}
with $\alpha=0,\dots,9$ denoting the different orbitals.
Given this, the LDOS is defined as
\begin{equation}
N^\alpha (\omega)=\sum_{\mathbf{k},\sigma}A^\alpha_\sigma(\mathbf{k},\omega).
\end{equation}
Figure \ref{fig:dos_hn} displays the LDOS for all orbitals, divided into three categories: [i] panels (a)-(b), for $(s0,s5)$ and $(s1,s8)$; [ii] panels (c)-(d), for $(s2,s7)$ and $(s3,s6)$; [iii] panels (e)-(f), for $(s4,s9)$.
Notice that this division is directly related to the coordination number $z$ of the orbitals.
The panels on the left [i.e., (a), (c), and (e)] display data for homogeneous hopping, $t_{\mathbf{i},\mathbf{j}}=1$, while those in the right [i.e., (b), (d), and (f)] have $t_{14}=t_{89} = 0.5$. Figure \ref{fig:dos_hn} clearly shows a well-formed van Hove singularity at $\omega=0$ in the LDOS of orbitals ($s0$,$s5$), ($s1$,$s8$), ($s2$,$s7$) and ($s3$,$s6$), even in the presence of anisotropy introduced in $t_{14}=t_{89}$.  Meanwhile, the peak at $\omega=0$ for $(s4,s9)$ is rather weak already for the isotropic case.

\begin{figure}[t] 
\centering
\includegraphics[width=0.98\linewidth]{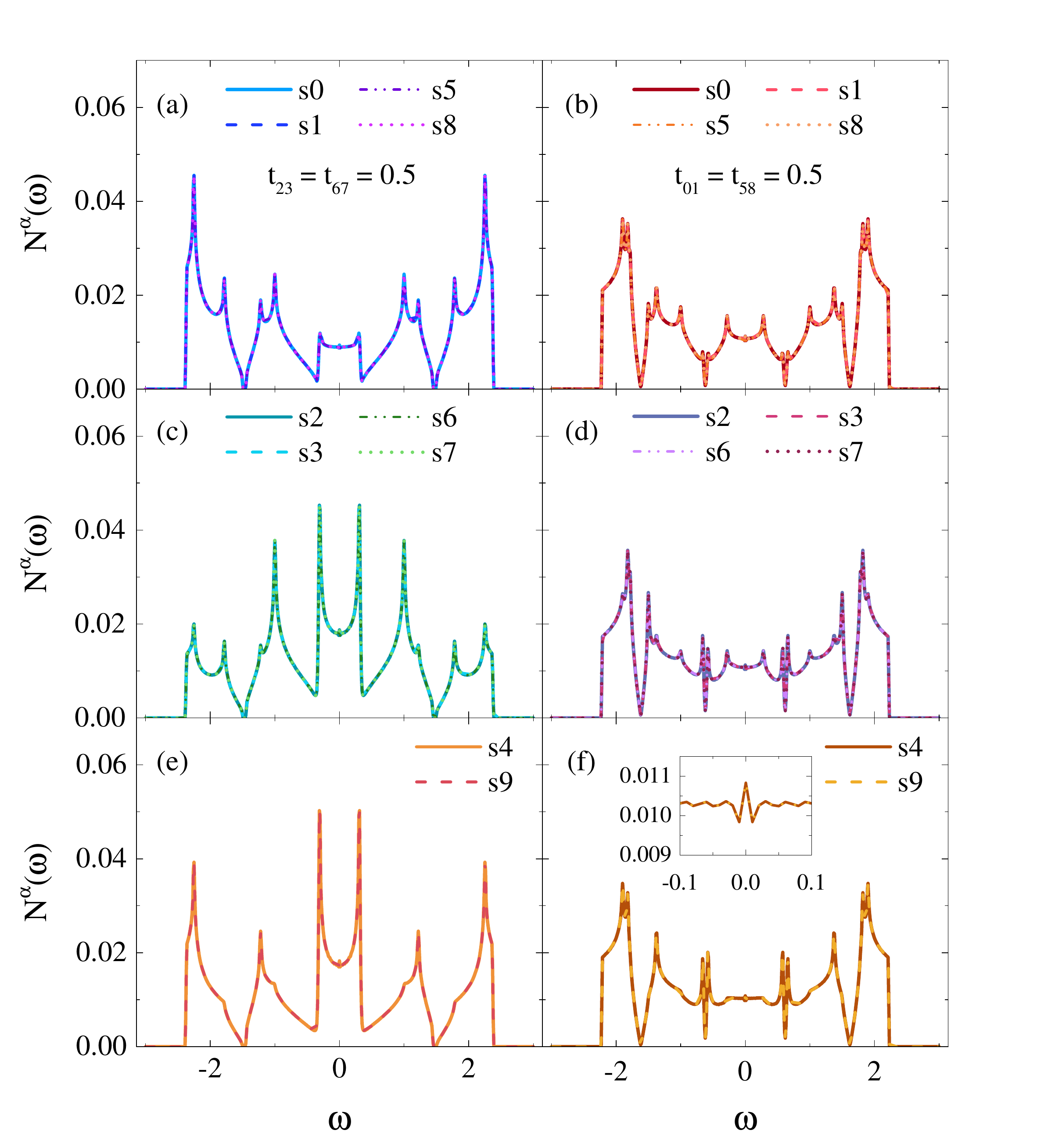}
\caption{Local density of states for orbitals $s0$, $s5$, $s1$ and $s8$ (a)-(b), $s2$, $s7$, $s3$ and $s6$ (c)-(d), and $s4$ and $s9$ (e)-(f). The panels on the left display the LDOS with $t_{23}=t_{67} = 0.5$ and the panels on the right  $t_{01}=t_{58} = 0.5$.
The LDOS becomes finite in panels (a)-(d) when hopping anisotropies $t_{23}=t_{67}$ and $t_{01}=t_{58}$ are introduced, resembling the behavior in Fig.\;\ref{fig:dos_hn}\;(e) and (f). The inset in panel (f) shows the LDOS for sites $s4$ and $s9$ near $\omega = 0$.}
\label{fig:DOSvar23670158}
\end{figure}

We can show that the van Hove singularity at the Fermi level is due to the crossing of the accidental and symmetry-enforced nodal lines at the momentum $\mathbf k=(0,\pi)$. Expanding the dispersion of the two low-energy bands in the vicinity of the crossing point, we obtain the form  $\varepsilon_\pm (\mathbf p)\approx \pm c |p_xp_y|$, where $\mathbf p={(k_x,k_y-\pi)}$ with $|\mathbf p|\ll 1$ and $c$ is a constant of the order of the bandwidth. We then have 
\begin{equation}
   \sum_\alpha N^\alpha(\omega)\sim \int d^2p \,\delta(\omega \mp c|p_xp_y|)\sim \ln\left(\frac{c\Lambda^2}{|\omega|}\right), 
\end{equation}
where $\Lambda$ is an ultraviolet momentum cutoff of order unity. Thus, the LDOS diverges logarithmically for $\omega\to 0$. As a consequence, the Stoner criterion strongly suggests that the system is unstable against interactions in this case. On the other hand, if glide-symmetry-preserving perturbations gap out the accidental nodal line, we expect the low-energy density of states to be finite and inversely proportional to the velocity of the linear dispersion around the remaining symmetry-enforced nodal line. To verify this statement, we consider a glide-symmetry-preserving anisotropy that acts on any pair of glide-related hopping parameters other than $t_{14}=t_{89}$ (by that excluding the special anisotropic case for which, as seen before, the accidental nodal line and the van Hove singularity at the Fermi level remain). In Fig.\,\ref{fig:DOSvar23670158}, we show the LDOS for two cases in which we vary a pair of hopping integrals, either ($t_{23}$,$t_{67}$) or   ($t_{01}$,$t_{58}$). In both cases, we can see a finite LDOS at the Fermi level, thus confirming that a glide-symmetry-preserving anisotropy that destroys the accidental nodal line also removes the van Hove singularity at the Fermi level.

\subsection{Mean-field approach}

We now discuss the mean-field solution of the interacting ($U/t\neq 0$) Hamiltonian in Eq.\,\eqref{Eq:Hamilt}.
Since the unit cell has multiple orbitals, we need to define average values for the local density and local magnetization, i.e.~$\langle n_{\mathbf{r}\alpha} \rangle = \langle c^{\dagger}_{\mathbf{r} \alpha \uparrow} c_{\mathbf{r} \alpha \uparrow} + c^{\dagger}_{\mathbf{r} \alpha \downarrow} c_{\mathbf{r} \alpha \downarrow} \rangle$ and $\langle m_{\mathbf{r}\alpha} \rangle = \frac{1}{2} \langle c^{\dagger}_{\mathbf{r} \alpha \uparrow} c_{\mathbf{r} \alpha \uparrow} - c^{\dagger}_{\mathbf{r} \alpha \downarrow} c_{\mathbf{r} \alpha \downarrow} \rangle$, respectively.
In addition, as the lattice is bipartite, at half-filling we expect the emergence of staggered magnetism with no charge order. Thus, in our mean-field approach we set $\langle n_{\mathbf{r}\alpha} \rangle = n = 1$, while $\langle m_{\mathbf{r}\alpha} \rangle = m_{\alpha}> 0$ for $\alpha$ even, and $\langle m_{\mathbf{r}\alpha} \rangle = m_{\alpha} < 0$ for $\alpha$ odd. 
As a result, the mean-field Hamiltonian reads
\begin{align}\label{Eq:Hamilt_MF}
 \nonumber \mathcal{H}_{\rm MF} & =  -t\sum_{\langle \textbf{r}\alpha,\textbf{r}^{\prime}\alpha^{\prime} \rangle,\sigma} \big( c_{\textbf{r}\alpha \sigma}^{\dagger}c_{\textbf{r}^{\prime}\alpha^{\prime} \sigma}+ {\rm H.c.} \big) - \mu \sum_{\textbf{r}\alpha, \sigma} n_{\textbf{r}\alpha\sigma}
\\  - & \; U \; \sum_{\textbf{r}\alpha, \sigma} \sigma \; m_{\alpha} \; c^{\dagger}_{\mathbf{r}\alpha\, \sigma} c_{\mathbf{r}\alpha\, \sigma} + N\,U\sum_{\alpha}\left(m_{\alpha}^{2} + \frac{1}{4}\right),
 \end{align}
with $\langle \textbf{r}\alpha,\textbf{r}^{\prime}\alpha^{\prime} \rangle$ denoting NN sites, and $N$ being the number of unit cells.

\begin{figure}[t]
\centering
\includegraphics[width=0.9\linewidth]{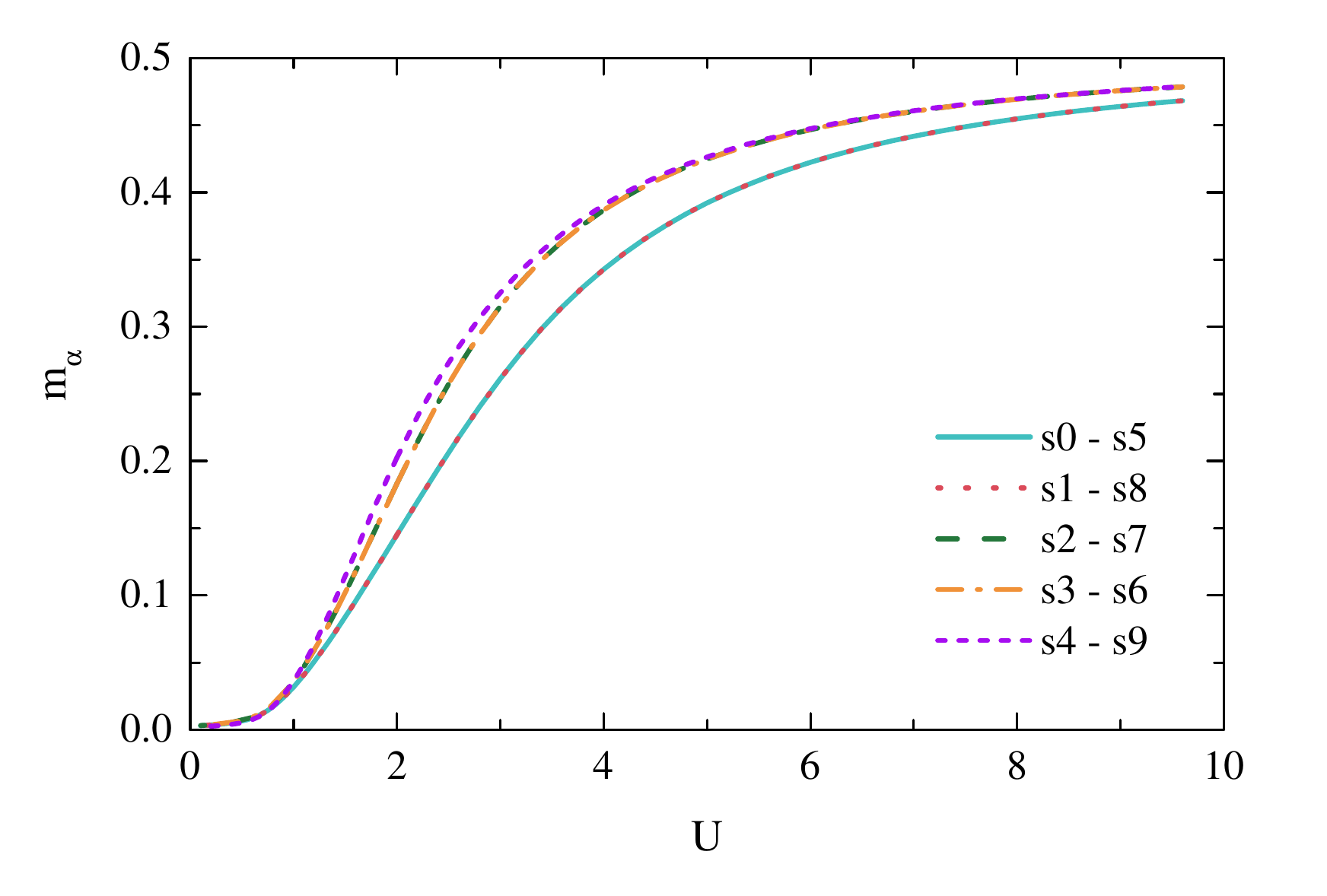} 
\caption{Magnetic order parameter as a function of the Hubbard interaction $U/t$.}
\label{fig:mean_field_afm}
\end{figure}

The mean-field Hamiltonian can be diagonalized in Fourier space using a ten-component spinor,  leading to the bands $E^{n}_{\mathbf{k},\sigma}$, with $n = 0,\ldots, 9$. The corresponding Helmholtz free energy is
\begin{equation}
\label{Helmholtz}
F = -\frac{1}{\beta} \sum_{n, \mathbf{k},\sigma} \ln \big( 1 + e^{-\beta E^{n}_{{\mathbf{k}},\sigma}} \big) + \text{const.} ,
\end{equation}
where $\beta=1/k_{B}T$.
We are then able to determine the effective fields by self-consistently minimizing the Helmholtz free energy, $\big\langle \frac{\partial F}{\partial m_{\alpha}} \big\rangle = 0$.
The resulting nonlinear coupled equations are solved numerically with standard library routine packages, using the Hellmann-Feynman theorem. Note that the term proportional to $m_\alpha$ in Eq.\,\eqref{Eq:Hamilt_MF} breaks the glide symmetry in the mean-field Hamiltonian by attributing an opposite magnetization to glide-related sites.
However, we assume that the magnetic phase preserves the product of glide symmetry and spin flip, which implies that pairs of sites connected by the glide symmetry must have opposite magnetization; for instance, $m_{0} = - m_{5}$, $m_{1} = - m_{8}$, and so on. This condition guarantees that the ground state has zero net magnetization and reduces the number of effective fields by half.

Figure \ref{fig:mean_field_afm} displays the ground state results for the mean-field magnetic order parameters $m_{\alpha}$ as a function of the interaction strength $U/t$, for homogeneous hopping integrals. Three remarks are in order. First, as expected, there are three distinct responses for the magnetic order parameters, due to the three kinds of LDOS discussed in the previous subsection. Note that the spontaneous magnetization is weaker on sites with coordination number $z=3$, corresponding to orbitals $(s0,s1,s5,s8)$, than on sites with only two nearest neighbors.  
Second, our mean-field approach shows magnetism for any $U/t>0$, putting the nonsymmorphic holey graphene in stark contrast to its pristine counterpart whose critical e-e interaction is known to be $U_{c}/t=3.8$\,\cite{Sorella12, Paiva05, Assaad2013}. The found magnetic order of nonsymmorphic holey graphene at arbitrarily weak e-e interaction is similar to the behavior of the Hubbard model on a square lattice, where the van Hove singularity is due to the nesting of the Fermi surface at half-filling, rather than the crossing of nodal lines as in our case.

As a third remark, we verified that the mean-field results are independent of the choice of spin direction used to break the \textit{SU(2)} symmetry. In particular, we obtained identical results when $\langle S_{\mathbf{i}}^{x} \rangle \neq 0$. Furthermore, we observed that any net staggered magnetization breaks the glide symmetry and is sufficient to open a gap at half-filling, while a ferromagnetic solution preserves the glide symmetry and, hence, the symmetry-enforced nodal line. However, it is important to emphasize that the ferromagnetic state is not the ground state of the system at this filling. We conclude that, although the lattice with staggered magnetization has a kind of ``modified glide symmetry", defined by the crystalline glide plane times a spin flip, this new symmetry does not ensure the survival of the nodal line enforced by the glide symmetry in the non-interacting model.



 \subsection{Linear spin wave theory}
 \label{subsec:lswt}
Since the mean-field solution discussed in the previous subsection is expected to be valid in the limit $U \to 0$, it is worthwhile to consider the strong-coupling limit, $U \to \infty$. At half-filling, the system can be mapped onto the Heisenberg model, whose ground-state magnetic properties can be analyzed using linear spin-wave theory (LSWT).

The Heisenberg model on the depleted honeycomb lattice is given by   \be
\mc H_{s} =\sum_{\langle \mb i,\mb j\rangle }J_{\mb i,\mb j}\mb S_{\mb i}\cdot \mb S_{\mb j},
\ee
where $J_{\mb i,\mb j}=4t^2_{\mb i,\mb j}/U$ is the exchange coupling and $\mb S_{\mb i}$ is the spin-$S$ operator at site $\mb i$.  We shall keep $S$ as a free parameter and set $S=1/2$ at the end. Assuming uniform hopping integrals, we set $J_{\mb i,\mb j}=J=4t^2/U$ for all nearest-neighbor bonds.

The classical spin configuration corresponds to a N\'eel state on this bipartite lattice.  To describe the spin excitations, we use the Holstein-Primakoff transformation with $S^z_{\mb i}=(-1)^\alpha (S-a^\dagger_{\mb i}a^{\phantom\dagger}_{\mb i})$, where $\alpha=0,1,\dots,9$ is the orbital index of site $\mb i$ and $a_{\mb i}$  is a boson annihilation operator. Within LSWT, the quadratic boson Hamiltonian to order $S$ has the form\be
\mc H_s\approx E_{\rm cl}+\frac{JS}2\sum_{\mb i}z_{\mb i} a^\dagger_{\mb i}a^{\phantom\dagger}_{\mb i}+JS\sum_{\langle \mb i,\mb j\rangle} (a^{\phantom\dagger}_{\mb i}a^{\phantom\dagger}_{\mb j}+a^\dagger_{\mb i}a^{ \dagger}_{\mb j}),\label{LSWH}
\ee
where $z_{\mb i}=z_\alpha$ is the coordination number of site $\mb i$ with orbital index $\alpha$  and $E_{\rm cl}=-12NJS^2$ is the energy of the classical N\'eel state for $N$ unit cells. 

Taking a Fourier transform, we define the vector $\mb X(\mb k)=(a^{\phantom\dagger}_{0}(\mb k),\dots,a^{\phantom\dagger}_{9}(\mb k),a_0^\dagger(-\mb k),\dots,a_9^\dagger(-\mb k))$ and rewrite the Hamiltonian in the form
\be
\mc H_s\approx E_{\rm cl}-NJS\text{Tr}(A)+\frac{JS}2\sum_{\mb k}\mb X^\dagger(\mb k) \mb H(\mb k) \mb X(\mb k),
\ee
where \be
\mb H(\mb k)=\left(\begin{array}{cc} A&B(\mb k)\\
B^*(-\mb k)&A
\end{array}\right).
\ee
Here $A$ is a $10\times 10$ diagonal matrix with nonzero elements given by  $A_{\alpha \alpha}=z_\alpha$ and $B(\mb k)$ stems from the Fourier transform of the nearest-neighbor-coupling terms in Eq.\,\eqref{LSWH}. We diagonalize the Hamiltonian in the standard way \cite{Chen2008} using a Bogoliubov transformation $\mb X(\mb k)=\mb Q(\mb k)\mb Y(\mb k)$, where $\mb Y(\mb k)$ is the new vector of bosonic operators and the matrix $\mb Q(\mb k)$  obeys \bea
\mb Q^\dagger(\mb k) \mb g \mb Q(\mb k)&=&\mb g,\\
\mb Q^\dagger(\mb k) \mb H(\mb k) \mb Q(\mb k)&=&\mb \Lambda(\mb k),
\eea
where $\mb g=\text{diag}(1,\dots,1,-1,\dots,-1)$ and $\mb \Lambda(\mb k)=\text{diag}(\omega_0(\mb k),\dots,\omega_{9}(\mb k),\omega_0(-\mb k),\dots,\omega_{9}(-\mb k))$. The functions $\omega_\lambda(\mb k)$, with $\lambda=0,\dots,9$, correspond to the dispersion relations of the magnon bands.  The quantum correction to the magnetization in each  sublattice   can be calculated from the  matrix $\mb Q(\mb k)$ as \be
\Delta S_\alpha= \frac{1}{2N}\sum_{\mb k}\sum_{\lambda=10}^{19}\left[|\mb Q_{\alpha,\lambda}(\mb k)|^2+|\mb Q_{\alpha+10,\lambda}(\mb k)|^2\right]-\frac12.
\ee

\begin{figure}[t]
        \includegraphics[width=0.98\columnwidth]{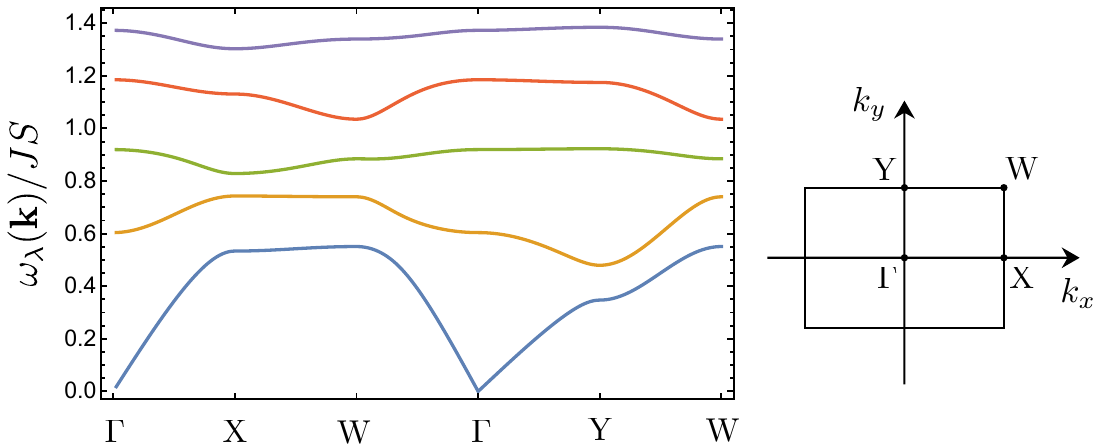}
    \caption{Dispersion relation of magnons for the Heisenberg model on the depleted honeycomb lattice.} \label{bands} 
\end{figure}

In Fig.\,\ref{bands} we show the magnon dispersion along high-symmetry directions in the Brillouin zone. We obtain five twofold degenerate bands. The lowest band is gapless with linear dispersion around the $\Gamma$ point, as expected from Goldstone modes associated with the spontaneous breaking of \textit{SU(2)} symmetry. The absence of zero-energy flat bands in the magnon spectrum indicates that the antiferromagnetic state is stable. Indeed, by calculating the quantum corrections to the magnetization,  we find that the order parameter remains finite and respects the symmetry that combines 
spatial inversion and a spin flip, as well as the composition of 
glide plane and spin flip.
The order parameters $m_\alpha=S-\Delta S_\alpha$ for each sublattice are  (setting $S=1/2$) \bea
m_0&=&m_1=m_5=m_8=0.124,\nonumber\\
m_2&=&m_3=m_6=m_7=0.140,\nonumber\\
m_4&=&m_9=0.199.
\eea
The average antiferromagnetic order parameter is $m=\frac1{10}\sum_\alpha m_\alpha=0.146$. Note that the order parameter is smaller   on the sites with coordination number $z=3$, in agreement with the mean-field result. Comparing with the result $m=0.242$ from LSWT for the pristine honeycomb lattice \cite{Reger1989}, we conclude that the creation of vacancies leads to weaker antiferromagnetic order in the strong-coupling limit. 
\begin{figure}[t]
\centering
\includegraphics[width=0.9\linewidth]{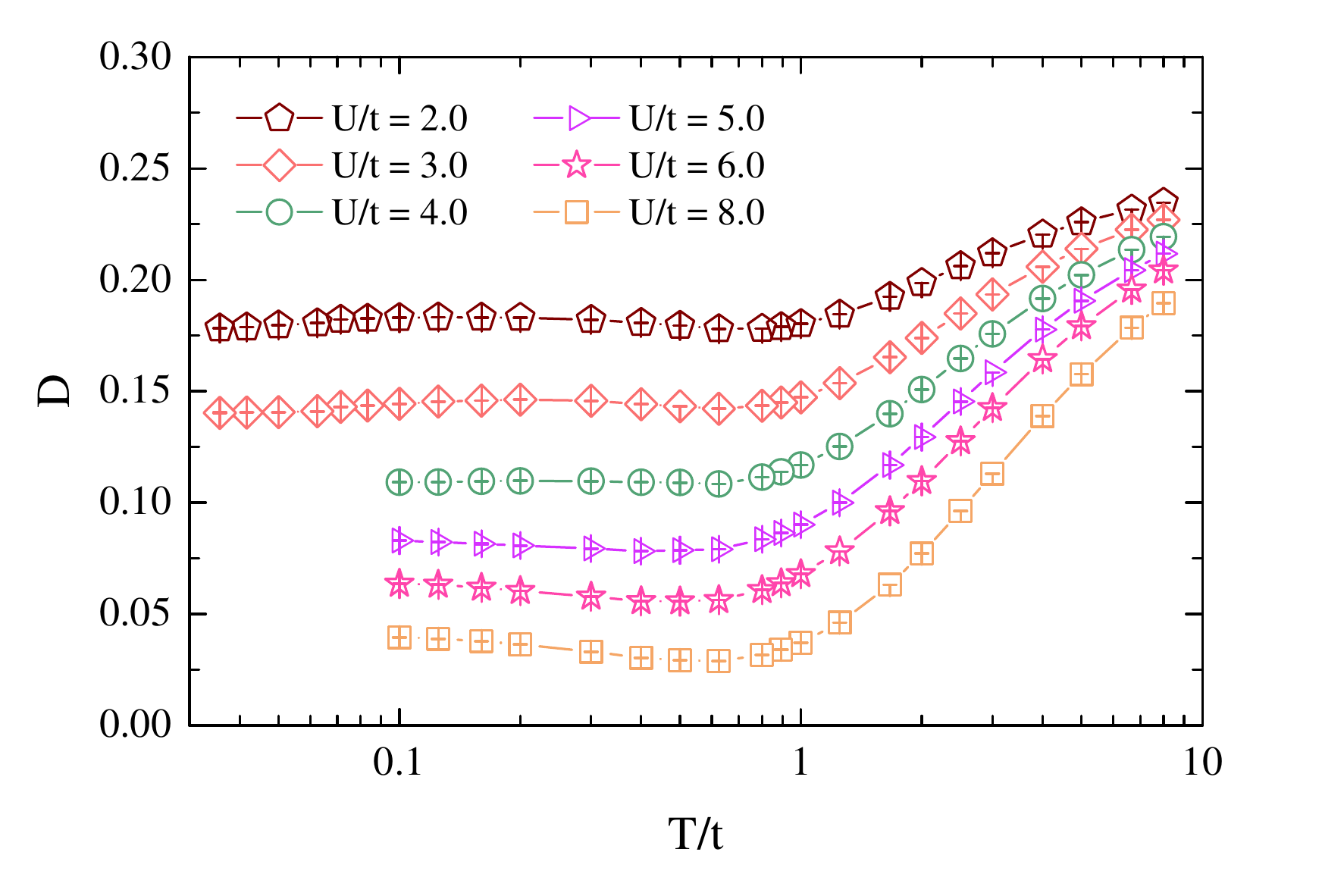} 
\caption{Double occupancy as function of $T/t$ for different $U/t$ considering a $N = 4 \times 4$ lattice (i.e.~160 sites).}
\label{fig:docc}
\end{figure}

 \subsection{Quantum Monte Carlo}

After discussing the weak- and strong-coupling limits in the previous subsections, we now turn our attention to the Hubbard model at intermediate interaction strengths. This analysis is performed using unbiased DQMC simulations, with up to $10^4$ Monte Carlo sweeps for thermalization and $10^5$ for measurements.
To this end, we start with the mean double occupancy,
\begin{equation}
D = \frac{1}{10 N} \sum_{\mathbf{r}, \alpha} \big\langle  n_{\mathbf{r},\alpha \uparrow} n_{\mathbf{r},\alpha \downarrow} \big\rangle = \frac{1}{10} \sum_{\alpha} D_{\alpha}~,
\end{equation}
which is the average over all orbitals, with $N$ being the number of unit cells.
Figure \ref{fig:docc} displays the behavior of $D$ as a function of temperature for different values of $U/t$.
The mean double occupancy is suppressed as $U/t$ increases, with the appearance of a minimum for sufficiently large interaction strength values.
As suppression of $D$ is expected for interacting systems (since electrons avoid sitting in the same orbital), its minimum is directly related to the emergence of unusual (bad) metallic properties \cite{Kim2020}.
In addition, there are many suggestions that the sign change of $\partial D / \partial T$ is a key feature of the occurrence of anomalies in the Seebeck coefficient, which, in turn, leads to a reconstruction of the Fermi surface\;\cite{Willdauany23,Nandini24}.
Finally, for $\frac{\partial  D}{\partial T} < 0$, the system can be adiabatically cooled by increasing the Hubbard interaction \cite{Werner05,Tremblay07,Paiva10,Paiva11,Paiva15,Medeiros23}, a condition relevant to the establishment of this geometry in cold-atom experiments. In summary, although beyond the scope of this paper, all of these features (bad metallicity, anomalous Seebeck effect, and adiabatic cooling) are likely to occur in our interacting NLSM.


The double occupancy is directly related to the local-moment formation, since $D_{\alpha} =  \frac{1}{2} \left[ \langle n_{\alpha} \rangle - \langle m^{2}_{\alpha} \rangle  \right]$, with $\langle m^{2}_{\alpha} \rangle = \langle (n_{\alpha \uparrow} - n_{ \alpha \downarrow})^2 \rangle$. Therefore, a decrease in $D_{\alpha}$ implies an increase in $\langle m^{2}_{\alpha} \rangle$, once $\langle n_{\alpha} \rangle = 1$ at half filling. Given this, the behavior of the orbital local moments $\langle m^{2}_{\alpha} \rangle$ as a function of temperature is similar to that presented in Fig.\,\ref{fig:docc}, for the mean double occupancy, but showing maxima instead of minima. 

\begin{figure}[t]
\centering
\includegraphics[width=0.9\linewidth]{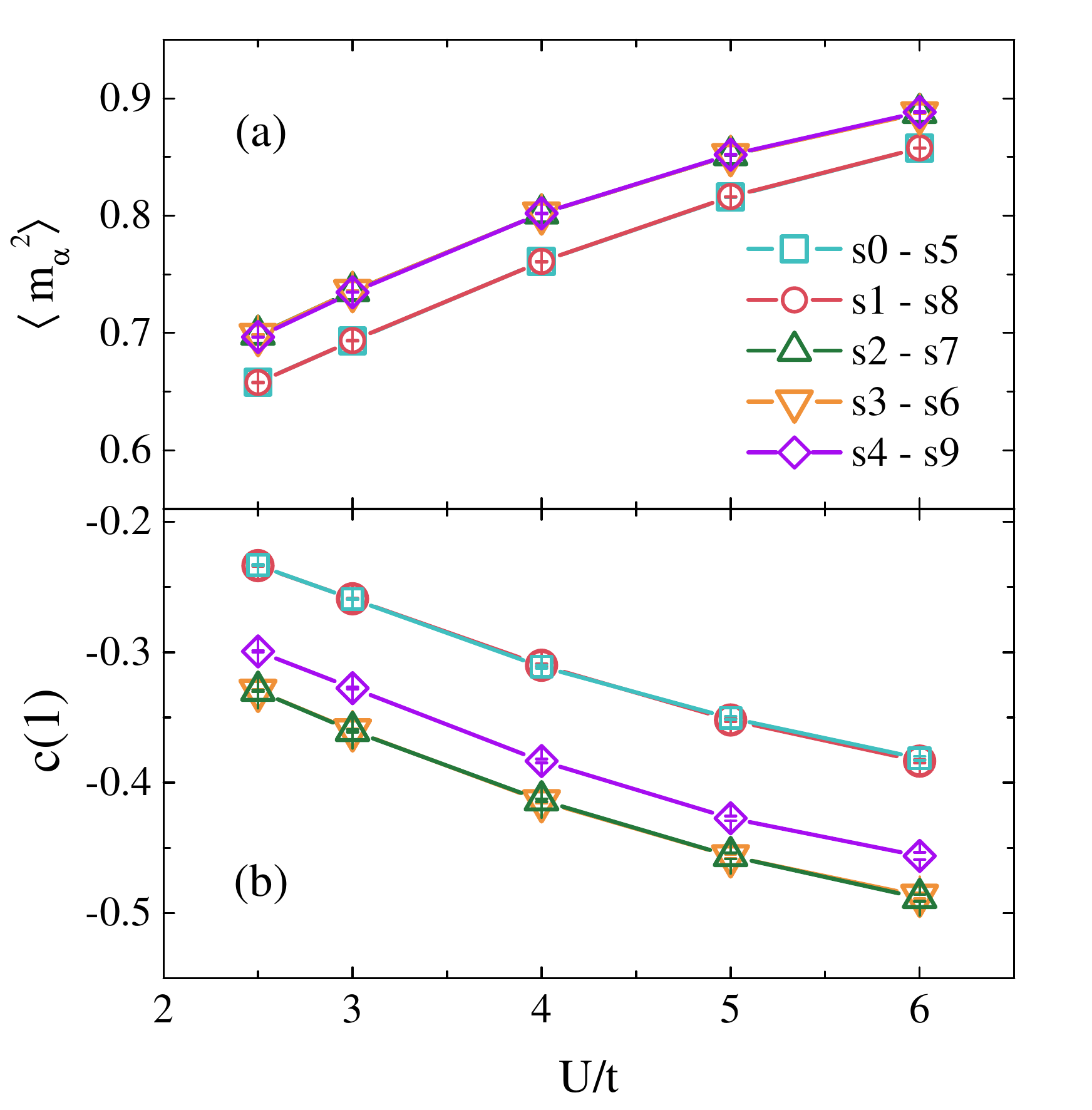}
\caption{(a) Local moment, (b) spin-spin correlation function for nearest neighbors at $\beta = 24$ as a function of $U/t$ for a depleted honeycomb lattice of size $N = 4 \times 4$ ($160$ sites).}
\label{fig:local_nn_nnn}
\end{figure}

To further investigate the magnetic properties, we examine the local moments as a function of $U/t$, at a low temperature ($\beta = 24$ is low enough to provide results close to the ground state ones), as displayed in Fig.\,\ref{fig:local_nn_nnn}\,(a). We notice that $\langle m^{2}_{\alpha} \rangle$ are well-formed and exhibit a non-uniform spatial response, with the local moment at orbitals $(s0,s5)$ and $(s1,s8)$ being smaller than at other ones. Longer distances spin-spin correlation functions, given by
$$c(\mathbf{i} - \mathbf{j}) = \langle (n_{\mathbf{i} \uparrow} - n_{\mathbf{i} \downarrow}) (n_{\mathbf{j} \uparrow} - n_{\mathbf{j} \downarrow}) \rangle = 4 \langle S^{z}_{\mathbf{i}} S^{z}_{\mathbf{j}} \rangle~,$$
may also be analyzed, where we keep the notation $\mathbf{i} = \boldsymbol\alpha + \mathbf{r}$.
Figure \ref{fig:local_nn_nnn}\,(b) illustrates the behavior of $c(\mathbf{i} - \mathbf{j})$ as a function of $U/t$ for each pair of orbitals corresponding to their NN sites, where $|\mathbf{i} - \mathbf{j}| = 1$. In fact, the quantity $c(1)$ represents the average value of $c(\mathbf{i} - \mathbf{j})$ over all NN sites.  
Notice that the non-uniform feature for the correlation functions remains, i.e.~orbitals $(s0,s5)$ and $(s1,s8)$ display weaker spin-spin correlations, in agreement with the mean-field results in Fig.\,\ref{fig:mean_field_afm}. 

\begin{figure}[t] 
\centering
\includegraphics[width=0.9\linewidth]{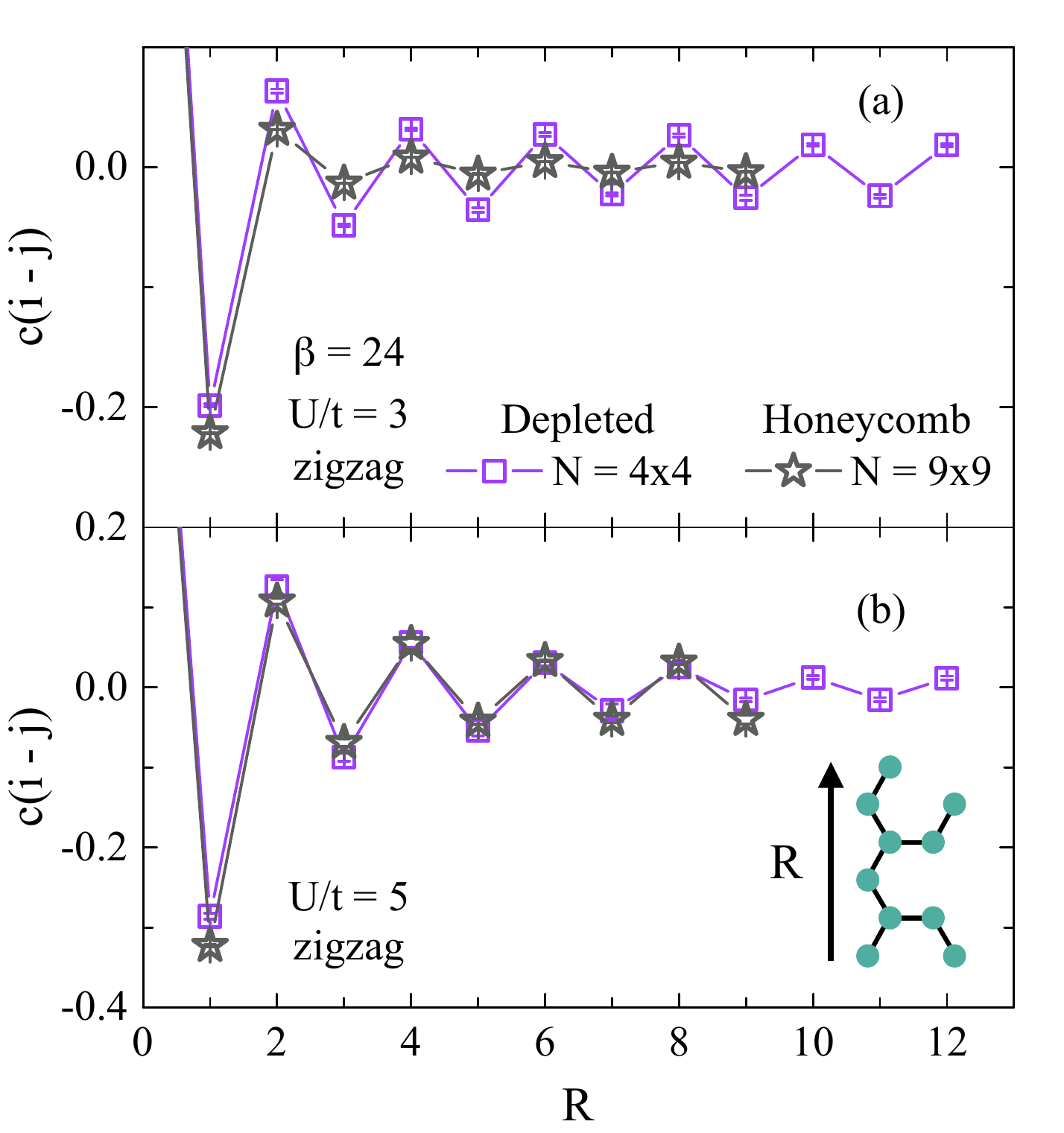}
\caption{Spatial dependence of the spin-spin correlation functions for (a)\;$U/t = 3$ and (b)\;$U/t = 5$ at $\beta = 24$ for a depleted honeycomb lattice of size $N = 4 \times 4$ (squares) and a pristine lattice of size $N = 9 \times 9$ (stars). The origin is set on site $s0$ and \textbf{r} runs along the sites in the zigzag direction (see inset). 
}
\label{fig:spincor_longdis}
\end{figure}

It is also worth comparing the magnetic properties of the depleted lattice with those of the pristine honeycomb one.
To this end, we analyze the spin-spin correlation functions between a reference orbital $s0$ and the \textbf{i}-th site/orbital along the zigzag direction. Specifically, following our definition of $\mathbf{i}\equiv (\mathbf{r},\alpha)$, the sites we examine are $(\mathbf{r},\alpha) = (\mathbf{0},0),~ (\mathbf{0},1),~ (\mathbf{0},4),~ (\mathbf{0},5),~ (\mathbf{0},8),~ (\mathbf{0},9),~ (\hat{y},0),~ (\hat{y},1), \dots$ (see, e.g., Fig.\,\ref{fig:honeydec}).
Figure \ref{fig:spincor_longdis} shows the spin-spin correlation functions $c(R_{\mathbf{i}})$, with $R_{\mathbf{i}}$ being the Hamming distance, for (a) $U/t=3$ and (b) $U/t=5$, while fixed $\beta=24$.
In the latter case, both the depleted and pristine honeycomb lattices display similar behavior. However,  for $U/t=3$, the $c^{\alpha,\gamma}(R_{\mathbf{i}})$ response of the depleted lattice is significantly stronger than that of the honeycomb lattice.
We recall that the Hubbard model in the honeycomb lattice has an AFM critical point at $U_{c}/t=3.8$, therefore the results of Fig.\,\ref{fig:spincor_longdis}\,(a) suggest that site depletion enhances magnetic correlations, at least for weak and intermediate interaction strengths.
This phenomenon has indeed been observed in the honeycomb lattice with a single depletion, and in our case, it suggests the emergence of magnetism below the critical $U/t$ of the pristine lattice\,\cite{Charlebois2015}.

The emergence of magnetism in the Hubbard model on depleted bipartite lattices is, in many cases, related to Lieb's theorem, which ensures the occurrence of ferrimagnetism when the number of sites in one sublattice exceeds that of the other sublattice\,\cite{Lieb89,Costa2016}. However, in the case we are considering, both sublattices have an equal number of sites. As a result, Lieb's theorem only ensures that the total spin is zero, leaving the possibility of magnetic ordering as an open issue. 

\begin{figure}[t]
\includegraphics[width=1.0\linewidth]{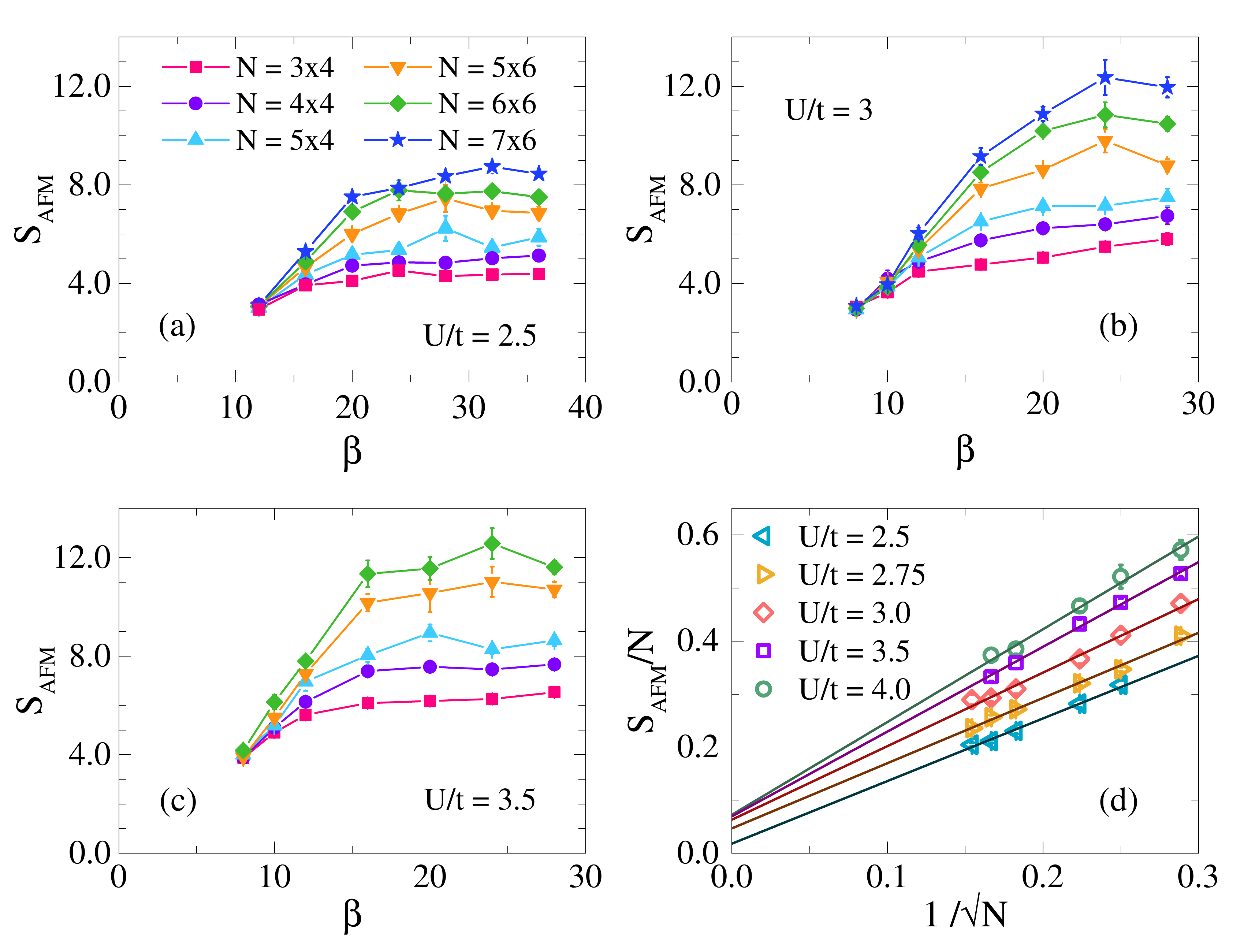} 
\caption{Antiferromagnetic structure factor as a function of $\beta$ for (a) $U/t = 2.5$, (b) $U/t = 3$ and (c) $U/t = 3.5$. 
At lower temperatures, the structure factor stabilizes, allowing for extrapolations to the thermodynamic limit\;\cite{Huse88}, as illustrated in panel (d). When not shown, error bars are smaller than symbol size.}
\label{fig:struct_fact}
\end{figure}

A thorough determination of long-range order is given by the AFM spin structure factor, 
\begin{align}\label{Eq:SAFM}
\nonumber S_{AFM} & = \frac{1}{10}\frac{1}{N}\sum_{\mathbf{i},\mathbf{j}} c(\mathbf{i} - \mathbf{j}) (-1)^{R_{ij}} \\
&= \frac{1}{10}\frac{1}{N}\sum_{\mathbf{i},\mathbf{j}} 4 \langle S^{z}_\mathbf{i} S^{z}_\mathbf{j} \rangle (-1)^{R_{ij}}~,    
\end{align}
which is proportional to the square of the order parameter in the thermodynamic limit.
Figure \ref{fig:struct_fact} shows $S_{AFM}$ as a function of the inverse of temperature, for fixed (a) $U/t=2.5$, (b) $U/t=3.0$ and (c) $U/t = 3.5$, and different system sizes $N$. In all cases, the following behavior is observed: at high temperatures (low $\beta$), $S_{AFM}$ is independent of the lattice size, indicating that the system exhibits only short-range correlations. As the temperature decreases (higher $\beta$), the spin correlation length becomes comparable to the system size, leading $S_{AFM}$ to develop a plateau, whose value depends on $N$. The lattice size dependence of such a plateau is an important hint of long-range order.
Once the structure factor no longer changes as the temperature is lowered, one can normalize the low-temperature results by the lattice size using the linear spin-wave scaling by Huse\,\cite{Huse88}, in which $$ S_{AFM}/N = A + B/\sqrt{N} + \mathcal{O}(1/N) $$  where $A= 4 m^{2}_{AFM}$, and with $m_{AFM}$ being the ground-state AFM order parameter in the thermodynamic limit\;\footnote{The constant 4 in $A = 4m^2_{AFM}$ arises from our definition of the structure factor $S_{AFM}$ in Eq.\,\eqref{Eq:SAFM}, which has the same multiplicative factor applied to $\langle S^z_{\mathbf{i}} S^z_{\mathbf{j}} \rangle$.}. 
This allows us to understand the behavior of the system at the thermodynamic limit ($1/N \rightarrow 0$ and $T = 0$ limit).
The extrapolations, shown in Fig.\,\ref{fig:struct_fact}\,(d), are finite in the $1/N \rightarrow 0$ limit, indicating that the system remains magnetic even for $U/t = 2.5$, in stark contrast to the pristine honeycomb lattice, which becomes magnetic only for $U/t \geq 3.8$\,\cite{Sorella12, Paiva05, Assaad2013}.

\begin{figure}[t]
\centering
\includegraphics[width=0.9\linewidth]{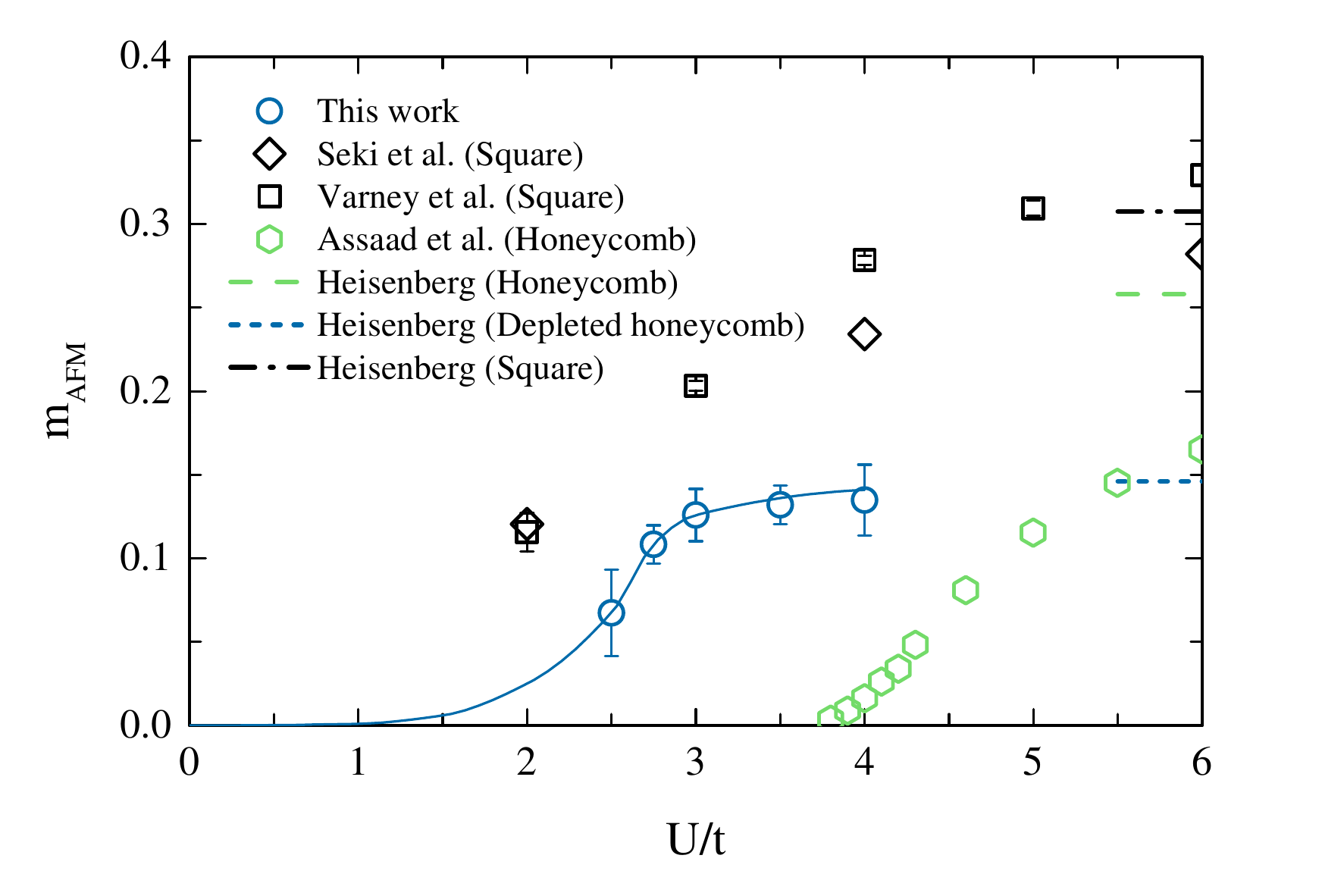} 
\caption{Antiferromagnetic order parameter as a function of $U/t$ (blue circle symbols). 
The points were obtained in the thermodynamic limit from the data displayed in Fig.\,\ref{fig:struct_fact}\,(d), as detailed in the main text. The solid line is a guide to the eye. The green hexagon symbols represent DQMC values for the honeycomb lattice (Ref.\,\cite{Assaad2013}), while the black square and diamond symbols are QMC results for the square lattice (Refs.\;\cite{Varney09} and \cite{Seki19} respectively). The horizontal lines denote the Heisenberg limit for the square (black dashed-dotted line; Ref.\,\cite{Sandvik2010}), pristine honeycomb (green dashed line; Ref.\,\cite{Castro2006}), and depleted honeycomb lattices (blue short dashed line; the value used here is the average AFM order parameter found in Sec.\;\ref{subsec:lswt}). }
\label{fig:order_param}
\end{figure}

The AFM order parameter $m_{AFM}$, obtained in Fig.\,\ref{fig:struct_fact}\,(d), is displayed in Fig.\,\ref{fig:order_param} (blue open circle symbols), along with the values of $m_{AFM}$ for the pristine honeycomb lattice presented in Ref.\,\cite{Assaad2013} (green open hexagon symbols). Notice that the strong-coupling limit for the depleted lattice (horizontal blue short-dashed line) is nearly achieved at $U/t = 4$. In contrast, larger values of $U/t$ are required to approach the strong coupling limit for the pristine honeycomb lattice (horizontal green dashed line). This difference in behavior can be attributed to instabilities associated with the van Hove singularity, which enhances the local moment formation as a function of $U/t$, effectively freezing the charge degrees of freedom. A similar feature is observed for the square lattice \cite{Varney09, Seki19}, as illustrated in Fig.\,\ref{fig:order_param}. However, the magnetization for the Heisenberg model on the depleted lattice is smaller than that of the square and pristine honeycomb lattices due to the reduced coordination number.
In light of these results, along with the MFT and LSWT results, we conclude that AFM order is likely present for any interaction strength in the depleted lattice.

\begin{figure}[t] 
\centering
\includegraphics[width=0.9\linewidth]{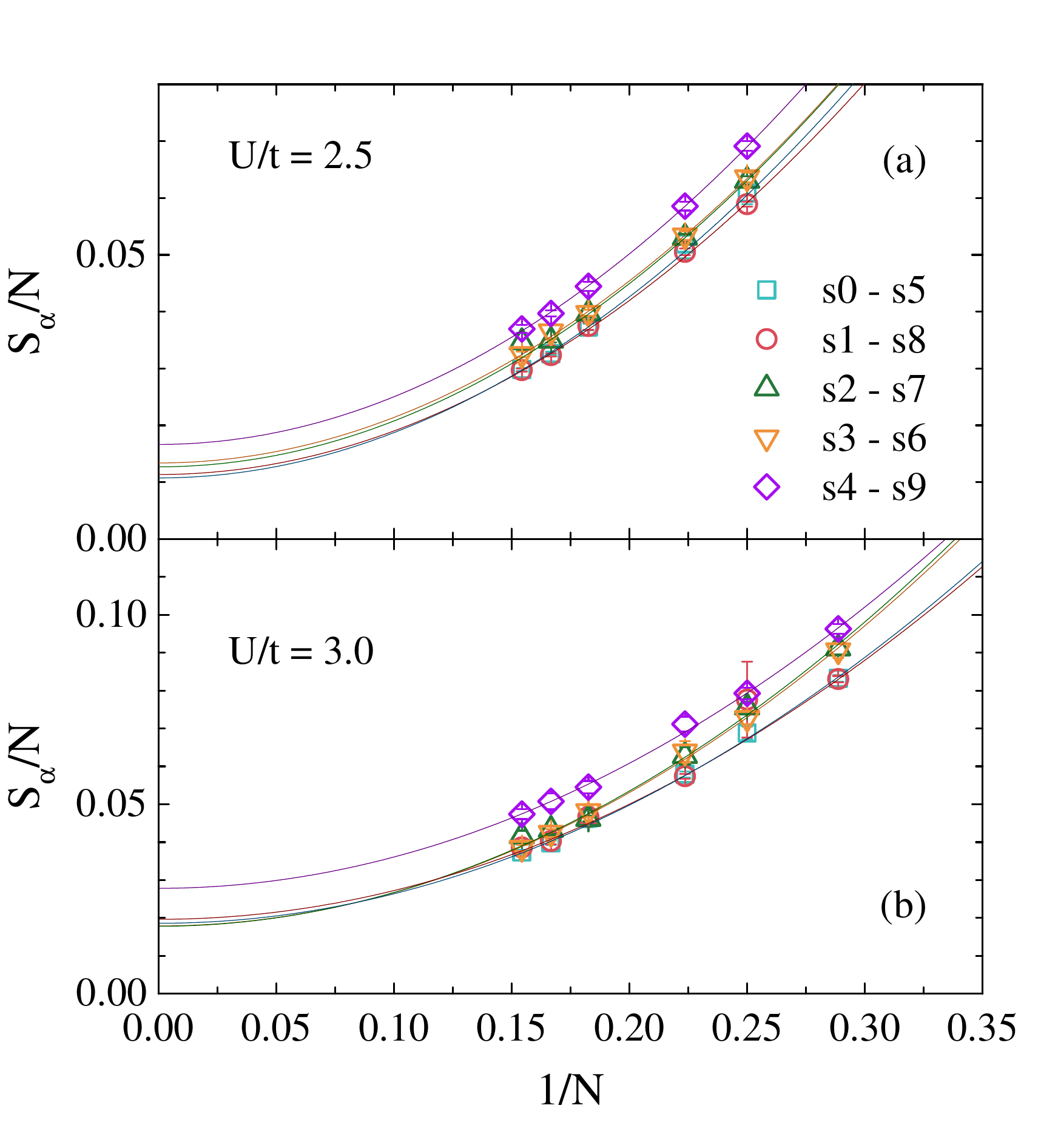}
\caption{Extrapolation of the  structure factor for individual orbitals as a function of inverse system size for (a) $U/t = 2.5$ and (b) $U/t = 3$.}
\label{fig:sFaa_pairs}
\end{figure}

We finish the analysis of the magnetic properties by examining the main orbital contributions to the AFM order. To this end, notice that Eq.\,\eqref{Eq:SAFM} may be written as  
\begin{equation}\label{Eq:SAFM2}
S_{AFM} = 4 \frac{1}{10}\frac{1}{N}\sum_{\mathbf{r},\mathbf{r}^{\prime}} \sum_{\alpha, \gamma} (-1)^{\alpha+\gamma}\langle S^{z}_{\mathbf{r}, \alpha}  S^{z}_{\mathbf{r}^{\prime}, \gamma} \rangle ~.
\end{equation}
Among the 55  possible orbital combinations, we focus on those where $\alpha = \gamma$, which represents the contribution of individual orbitals to the global order parameter. We then define
\begin{equation}\label{Eq:SF_individuals}
S_{\alpha} = \frac{1}{N}\sum_{\mathbf{r},\mathbf{r}^{\prime}} \langle S^{z}_{\mathbf{r}, \alpha}  S^{z}_{\mathbf{r}^{\prime}, \alpha} \rangle.
\end{equation}
Notice that the spin-spin correlations for orbitals on the same sublattice are ferromagnetic; thus, $S_{\alpha}>0$.  Figure \ref{fig:sFaa_pairs} presents the behavior of  $S_{\alpha}$, averaged over equivalent orbitals,  for (a) $U/t = 2.5$ and (b) $U/t=3$,  as a function of the system size at low temperature (i.e.~for large values of $\beta$, where the structure factor has stabilized as in Fig.\,\ref{fig:struct_fact}). An important remark must be made here: There is a noticeable difference between the contributions from the orbitals $(s4, s9)$ and the others.
Interestingly, these orbitals have the highest responses at the mean-field solution and in the LSWT approach.

\begin{figure}[t]
\centering
\includegraphics[width=0.9\linewidth]{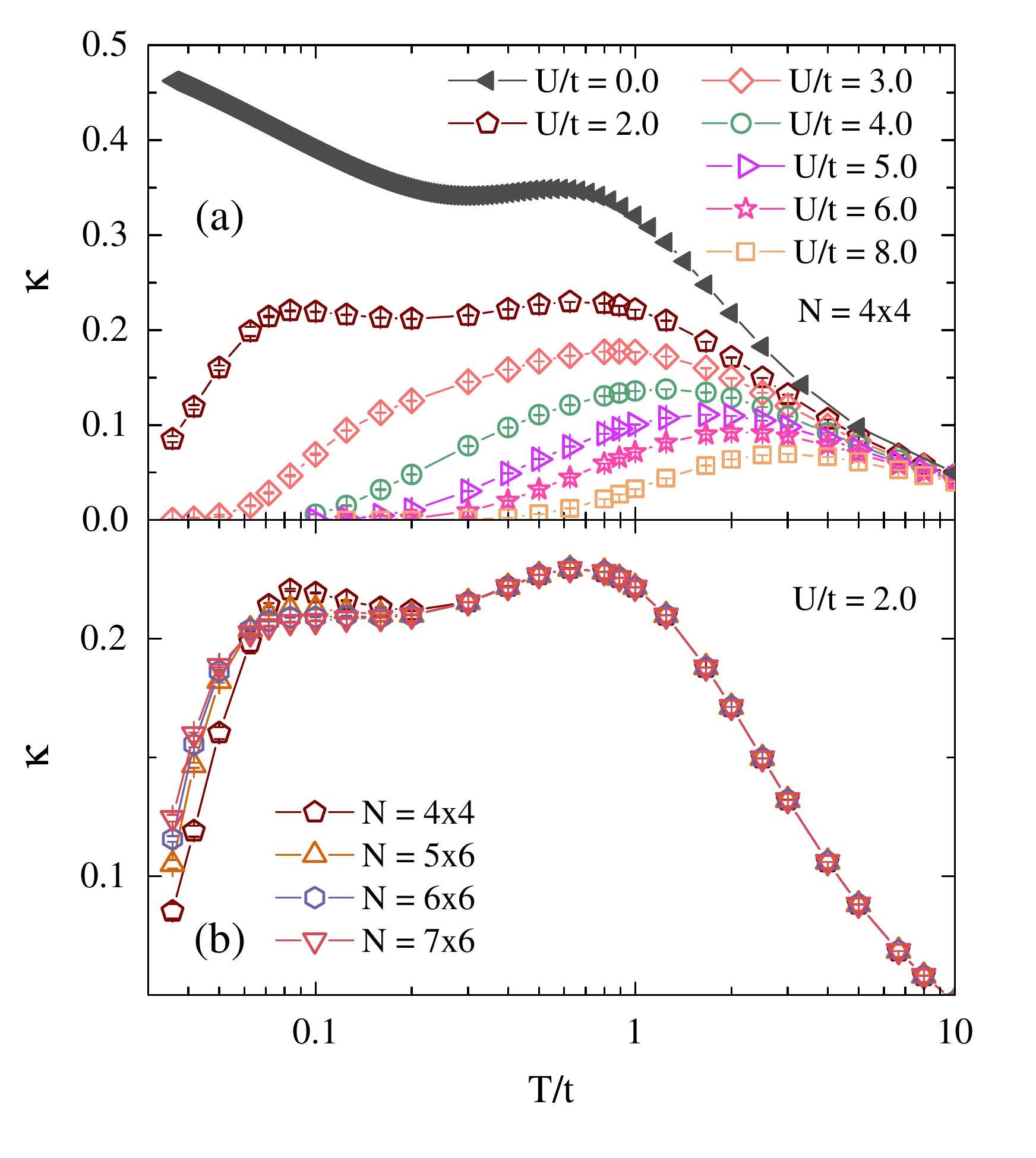} 
\caption{(a)\;Electronic compressibility as a function of $T/t$ for different values of $U/t$ and a depleted honeycomb lattice of size $N = 4 \times 4$. (b)\;Comparison of $\kappa$ for different lattice sizes and $U/t = 2.0$. For $U/t=2$, the compressibility displays a double-peaked structure due to finite size effects, as can be seen in the $N = 4 \times 4$ curve. A careful analysis with system size shows that the peak at $T/t \approx 0.083$ disappears as the system size increases. Therefore, the peak at $T/t = 0.625$ is the actual low-temperature peak for $U/t = 2.0$.}
\label{fig:compress}
\end{figure}

\begin{figure}[t]
    \centering
    \includegraphics[width=0.9\linewidth]{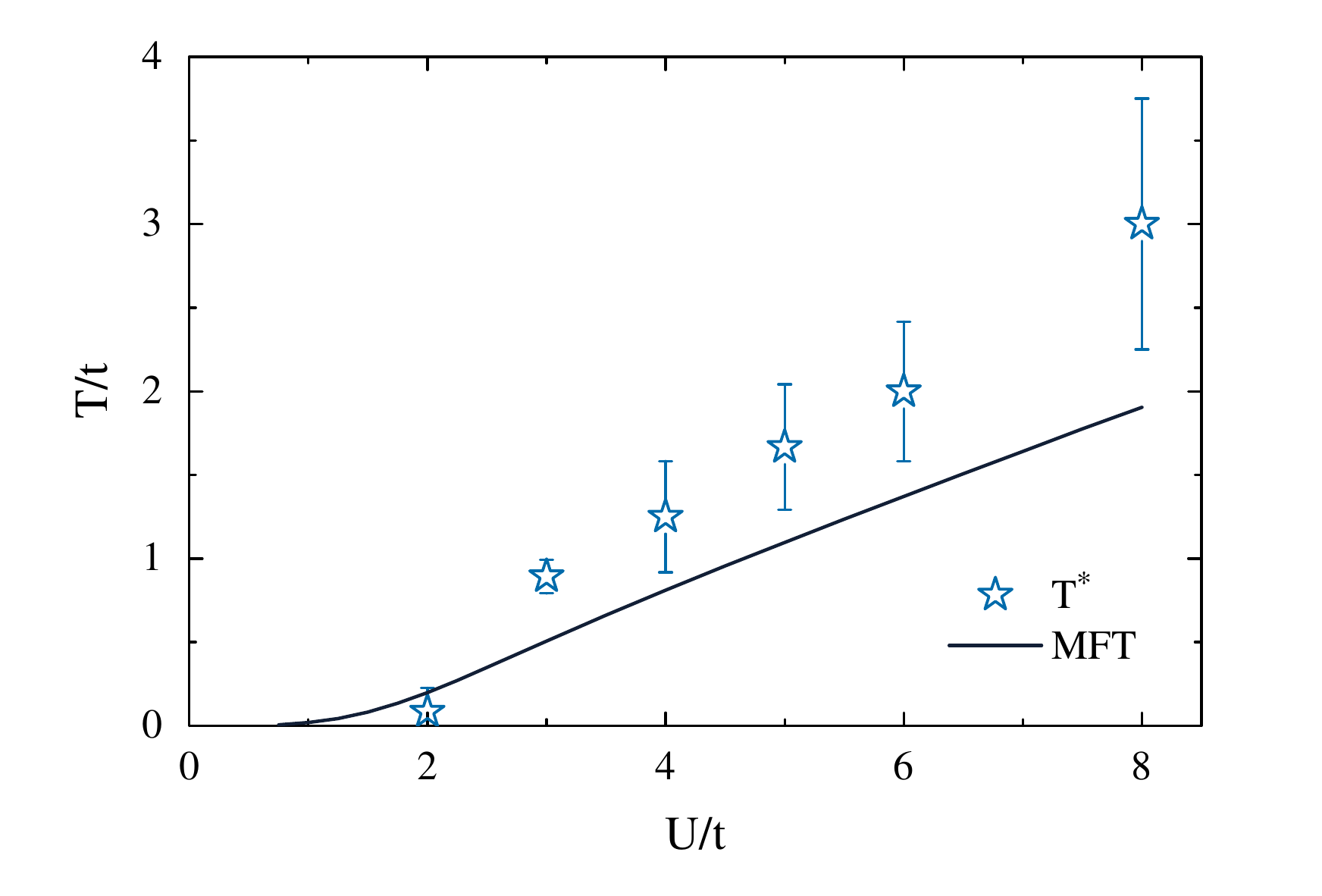}
    \caption{Temperature scale $T^*$ defined by the low-temperature maxima for the compressibility (open stars) and mean-field N\'eel temperature (solid line).}
    \label{fig:temp_scale}
\end{figure}

Now, we turn our attention to the electronic compressibility $\kappa = \frac{1}{n^{2}} \frac{\partial n}{\partial\mu}$, with $n = \frac{1}{N} \big\langle  \sum_{\mathbf{i}, \sigma} n_{i \sigma} \big\rangle $.
In order to perform the derivative $\frac{\partial n}{\partial\mu}$, we vary the chemical potential $\mu$ by $\Delta \mu = 0.05$, thus changing the density $n$.
The result is displayed in Fig.\,\ref{fig:compress}\,(a) as a function of temperature. In a metallic phase, $\kappa$ has finite values, but for an insulator we have $\kappa \to 0$ because $n(\mu)$ exhibits a plateau as $T \to 0$. 
These features are present in our interacting NLSM, as displayed in Fig.\,\ref{fig:compress}\,(a), with $\kappa \to 0$ for any $U/t>0$ at low temperatures. That is, as the temperature is lowered, the system enters an insulating phase with strong short-range spin correlations, which eventually give rise to long-range AFM order in the ground state.
It is important to note that the curve for $U/t = 2.0$ in Fig.\,\ref{fig:compress}\,(a) has two peaks, the first at $T/t = 0.625$ and the second at $T/t = 0.083$. 
However, in Fig.\,\ref{fig:compress}\,(b) it is possible to see that $\kappa$ depends on the system size for temperatures lower than $T/t < 0.2$, implying that the peak at $T/t = 0.083$ is actually a finite size effect and should disappear as $N \to \infty$.
Therefore, in Fig.\,\ref{fig:temp_scale} we have considered the peak at $T/t = 0.625$.

The temperature scale for the onset of the insulating phase, denoted as $T^{*}$, is presented in Fig.\,\ref{fig:temp_scale}.  It corresponds to the temperature of the lowest peak in the compressibility. Within a mean-field approach, long-range order occurs at the same temperature as the metal-insulator transition. Therefore, it is important to compare the mean-field N\'eel temperature with the $T^*$ from QMC, as presented in Fig.\,\ref{fig:temp_scale}. Interestingly, these energy scales are similar, indicating that the mean-field approach successfully captures short-range effects.

\section{Conclusions}
\label{sec:conc}


In this work, we investigate the interacting properties of the repulsive Hubbard model on a nonsymmorphic depleted honeycomb lattice which, in the noninteracting regime, exhibits symmetry-enforced and accidental nodal lines for the isotropic lattice and for a special anisotropic one. The crossing of the nodal lines leads to a van Hove singularity at zero energy in the local density of states, making the system unstable to interactions at half-filling. Generic anisotropy gaps the accidental nodal line, thus removing the associated van Hove singularity.

In the interacting case ($U/t>0$), we analyze the model using three different approaches: mean-field theory (MFT), linear spin wave theory (LSWT), and determinant quantum Monte Carlo (DQMC) simulations.
This allows us to examine the weak, strong, and intermediate coupling limits, respectively.

Within the mean-field approach, we observe that the system becomes magnetic and opens a gap in the spectrum for any $U/t > 0$, which can be attributed to the presence of the van Hove singularity in the noninteracting regime. Moreover, the magnetism is inhomogeneous within the unit cell, with orbitals $s4$ and  $s9$ exhibiting the largest responses. The onset of an inhomogeneous magnetic order for arbitrarily weak \textit{e}-\textit{e} interaction in nonsymmorphic holey graphene is to be contrasted with the magnetization of the pristine honeycomb lattice, which is homogeneous and requires $U/t > 3.8$.

In the opposite limit, as $U/t\to \infty$, our LSWT analysis for the Heisenberg model yields a global antiferromagnetic order parameter $m_{AFM} \approx 0.146$, which is significantly smaller than the value for the pristine honeycomb lattice, $m_{AFM} \approx 0.258$\,\cite{Castro2006}. This reduction is attributed to the smaller effective coordination number in the former case. Nevertheless, in agreement with the previous MFT solution, the magnetic responses exhibit inhomogeneities, with the largest contributions again arising from orbitals $s4$ and $s9$.

For the intermediary regime of \textit{e}-\textit{e} interactions, we employed QMC simulations  to investigate the isotropic lattice. We started by analyzing double occupancy, which exhibits a change in the sign of $\partial D/ \partial T$ as a function of temperature. This behavior is crucial for identifying bad metalicity, anomalies in the Seebeck coefficient, and adiabatic cooling. Additionally, our analysis of the spin-spin correlation functions reveals an inhomogeneous magnetic response, consistent with both the MFT and LSWT limits, and stronger correlations than that of the pristine honeycomb lattice in the weakly interacting regime.  Consistent with the MFT result, we obtain a staggered antiferromagnetic order for any $U/t>0$ also in the intermediary \textit{e}-\textit{e} interaction regime, a stark contrast to the Hubbard model on the honeycomb lattice.
It is important to notice that the MFT results are valid in the $U \to 0$ limit, where the antiferromagnetic order parameter becomes exponentially small. Outside this limit, MFT overestimates the order parameter, which renders a quantitative comparison of our DQMC and MFT results impossible.
For large values of $U/t$, the QMC order parameter agrees with the LSWT results, valid in the strongly interacting regime.
Finally, our analysis of the electronic compressibility shows insulating behavior at low temperatures with long-range antiferromagnetic order and metal-insulator transition occurring simultaneously. This collection of results demonstrates that vacancy engineering can qualitatively change the magnetic phase transition of a crystal and substantially alter the magnetization and spin correlations in the magnetic phase, thus providing a practical mechanism for designing quantum materials with tailored magnetic characteristics.

\section*{ACKNOWLEDGMENTS}
We are grateful to N. C. Costa for his contributions to this work. The authors are grateful to the Brazilian agencies Conselho Nacional de Desenvolvimento Cient\'\i fico e Tecnol\'ogico (CNPq) and Coordena\c c\~ao de Aperfei\c coamento de Pessoal de Ensino Superior (CAPES).
M.M. acknowledges financial support from FAPDF grant number 00193-00001817/2023-43. R.G.P. acknowledges support by a grant from the Simons Foundation (Grant No. 1023171) and by Finep (Grant No. 1699/24 IIF-FINEP). T.P. acknowledges financial support from Funda\c{c}\~ao Carlos Chagas Filho de Amparo \`a Pesquisa do Estado do Rio de Janeiro grant numbers E-26/200.959/2022  and E-26/210.100/2023; and from CNPq grant numbers 308335/2019-8, 403130/2021-2,  and 442072/2023-6; and also Instituto Nacional de Ci\^encia e Tecnologia de Informa\c c\~ao Qu\^antica (INCT-IQ).


\bibliography{honeydep}
\end{document}